\documentclass[prb,twocolumn,amssymb,amsmath,floatfix,showpacs]{revtex4}
\usepackage{bm,dcolumn,graphicx,hyperref}
\newcommand{\vect}[1]{\mathbf{#1}}
\newcommand{\unit}[1]{\hat{\mathbf{#1}}}
\newcommand{\kp}{\ensuremath{\vect{k} \cdot \vect{p}}}

\DeclareMathOperator{\im}{Im}
\begin{document}
\title{Choosing a basis that eliminates spurious solutions in
       $\mathbf{k} \cdot \mathbf{p}$ theory}
\author{Bradley A. Foreman}
\email{phbaf@ust.hk}
\affiliation{Department of Physics,
             Hong Kong University of Science and Technology,
             Clear Water Bay, Kowloon, Hong~Kong, China}

\begin{abstract}
A small change of basis in $\mathbf{k} \cdot \mathbf{p}$ theory yields
a Kane-like Hamiltonian for the conduction and valence bands of
narrow-gap semiconductors that has no spurious solutions, yet provides
an accurate fit to all effective masses.  The theory is shown to work
in superlattices by direct comparison with first-principles
density-functional calculations of the valence subband structure.  A
reinterpretation of the standard data-fitting procedures used in
$\mathbf{k} \cdot \mathbf{p}$ theory is also proposed.
\end{abstract}

\pacs{73.21.-b, 73.61.Ey, 71.15.Ap, 71.20.Nr}

\maketitle

\section{Introduction}

The Kane model for coupled conduction and valence electrons in
narrow-gap bulk semiconductors \cite{Kane57,Kane66,Kane80} was first
applied to superlattices three decades ago. \cite{SaiHalasz77} Today
this model is still used frequently for the study of medium- and
narrow-gap nanostructures.
\cite{Mlinar05,Novik05,Alfthan05,Kopr05,MaslNing05,Nil06,Box06,Wu06,%
Lassen06} Kane's theory has a notorious pitfall: spurious solutions
with large crystal momentum $\vect{k}$, which arise from small
Hamiltonian matrix elements of order $k^2$.
\cite{WhSh81,WhMaSh82,ScTH85,SmMa86,EpScCo87,RamMohan88,Trze88b,%
Bast88,SmMa90} \\ \cite{MaSm90a,Bast91,WiRo93,AvSi94,MeGoOR94,Duff94,%
SzmuBr95,GoMa96,Szmu96,Fore97,note:Fore97,Burt98a,KiGeLu98,Efros98,%
Burt99,Sercel99,Wang00,Jask01,Rod02,Holm02,Cart03,Kolokolov03,%
Szmu05a,Szmu05b} \\
\cite{Alfthan05,Kopr05,MaslNing05,Yang05,Box06,Wu06,Lassen06} 
Spurious propagating waves pose a serious problem, since their
presence within the energy gap changes the physical character of the
model system from semiconducting to metallic.

Many schemes for eliminating the unphysical effects of spurious
solutions have been proposed (e.g., changing or adding parameters in
the Hamiltonian, or excising the offending modes numerically or
analytically), but none has yet found wide acceptance.  The relative
merits of the various proposals are not discussed here.  Instead, it
is merely noted that all of these schemes take the form of patches
applied to Kane's original \kp\ theory.  The possibility of
reconstructing \kp\ theory on a different foundation has not been
considered.

This paper derives from first principles an $8 \times 8$ \kp\
Hamiltonian with no spurious solutions.  The key step is a slight
change in the standard choice of basis.  This allows the
adjustment-of-parameters method of Ref.\ \onlinecite{Fore97}, which
was proposed only as a useful approximation, to be formulated
rigorously.  The present derivation proves that---within the
limitations imposed by a second-order differential equation---{\em
this method is not an approximation}.  That is, all terms of order
$k^2$ derived from a clearly defined basis can be included without
approximation.  (The number of fitting parameters can be reduced with
a few standard approximations, \cite{Kane66,Fore97,PiBr66} but that is
not a fundamental limitation of the method.)  The change of basis is
applied here to the first-principles envelope-function theory
developed in Refs.\ \onlinecite{Fore05b}, \onlinecite{Fore07a}, and
\onlinecite{Fore07b}.  A comparison with density-functional
calculations on In$_{0.53}$Ga$_{0.47}$As/InP superlattices shows very
good agreement.

In conventional \kp\ perturbation theory, \cite{LuKo55,BirPik74} one
uses a unitary transformation to construct a basis in which the \kp\
coupling between the states of interest (set $\mathcal{A}$) and all
other states (set $\mathcal{B}$) is reduced to zero, while
simultaneously renormalizing the masses in $\mathcal{A}$ and
$\mathcal{B}$.  If $\mathcal{A}$ includes the highest valence and
lowest conduction states, the \kp\ coupling within $\mathcal{A}$ is
either set to zero (in single-band effective-mass theory
\cite{Kane80,LuKo55}) or not modified at all (in the multiband Kane
theory \cite{Kane66,Kane80,BirPik74}).

In the present approach, a unitary transformation is used to modify
the conduction--valence \kp\ interaction by only a small amount.  The
coupling can be either strengthened or weakened; its actual value is
fixed (in one of several possible choices) by setting the partially
renormalized conduction-band mass to zero.  This is precisely the
method used to eliminate spurious solutions in Ref.\
\onlinecite{Fore97}.  However, the interface operator ordering derived
here is more subtle than the simple heuristic model of Ref.\
\onlinecite{Fore97}.  The present theory also suggests the need for a
reinterpretation of the standard data-fitting procedures used in \kp\
models.

The situation encountered here is analogous to a gauge transformation
in quantum electrodynamics.  Although all gauges are equivalent in
exact calculations, different gauges may yield different predictions
in approximate calculations. \cite{CohTan89} Likewise, the unitary
transformation defined here would have no effect in an exact
calculation, but in a second-order \kp\ Hamiltonian of finite
dimension, varying the parameters of the unitary transformation
generates a metal-insulator phase transition in the model system.  The
remedy proposed here is simply to choose transformation parameters
that lie within the physical (i.e., insulating) regime of the phase
diagram.

The paper begins in Sec.\ \ref{sec:bulk} with the definition and
application of the unitary transformation to bulk semiconductors.  The
theory is extended to heterostructures in Sec.\
\ref{sec:heterostructures} and applied to the widely used
Pidgeon--Brown Hamiltonian \cite{PiBr66} in Sec.\ \ref{sec:PB}.
Numerical applications of the theory are presented in Sec.\
\ref{sec:numerical}.  Finally, the results of the paper are summarized
and discussed in Sec.\ \ref{sec:conclusions}.

\section{Bulk crystals}

\label{sec:bulk}

\subsection{Hamiltonian}

Consider first the case of a bulk semiconductor.  It is assumed at the
outset that a Luttinger--Kohn (LK) unitary transformation
\cite{LuKo55,BirPik74} has already been used to eliminate the \kp\
coupling between sets $\mathcal{A}$ and $\mathcal{B}$.  Thus, the
effective Hamiltonian $H$ for states in $\mathcal{A}$ is (in the LK
basis)
\begin{subequations} \label{eq:H}
  \begin{gather}
    \langle n \vect{k} | H | n' \vect{k}' \rangle = H_{nn'} (\vect{k})
    \delta_{\vect{k}\vect{k}'} , \label{eq:Hnknk} \\
    H_{nn'} (\vect{k}) = E_n \delta_{nn'} + k_{i} \pi^{i}_{nn'} +
    k_{i} k_{j} D^{ij}_{nn'} , \label{eq:Hnnk}
  \end{gather}
\end{subequations}
in which $E_n$ is the energy of state $n$ at $\vect{k} = \vect{0}$,
$\pi^{i}_{nn'}$ is the $i$ component of the kinetic momentum matrix,
and $D^{ij}_{nn'}$ is the inverse effective-mass tensor (in atomic
units)
\begin{multline}
  D^{ij}_{nn'} = \frac12 (\delta_{ij} \delta_{nn'} + i \epsilon_{ijk}
  \sigma^k_{nn'}) \\ + \frac12 \sum_{l}^{\mathcal{B}} \biggl(
  \frac{\pi^{i}_{nl} \pi^{j}_{ln'}}{\omega_{nl}} + \frac{\pi^{i}_{nl}
  \pi^{j}_{ln'}}{\omega_{n' l}} \biggr) , \label{eq:D}
\end{multline}
where $\omega_{nl} = E_n - E_l$.  Here and below all equations are
written (for simplicity) as if the potential energy were local,
although a nonlocal pseudopotential was used for the numerical
calculations in Sec.\ \ref{sec:numerical}.  The term
$\sigma^{k}_{nn'}$ is a matrix element of the Pauli spin operator,
which accounts for the intrinsic magnetic dipole moment of the
electron. \cite{Sak67_p78}

In Ref.\ \onlinecite{Fore97}, it was assumed to be permissible to
treat $\pi^{i}_{nn'}$ as an adjustable parameter in Eq.\ (\ref{eq:H}).
In this approach, $\pi^{i}_{nn'}$ is replaced by $\bar{\pi}^{i}_{nn'}
= \pi^{i}_{nn'} + \Delta \pi^{i}_{nn'}$, in which $\Delta
\pi^{i}_{nn'}$ has the same symmetry as $\pi^{i}_{nn'}$ and vanishes
when $E_n = E_{n'}$, but is otherwise arbitrary.  The Hamiltonian
(\ref{eq:Hnnk}) is then replaced by
\begin{equation}
  \bar{H}_{nn'} (\vect{k}) = E_n \delta_{nn'} + k_{i}
  \bar{\pi}^{i}_{nn'} + k_{i} k_{j} \bar{D}^{ij}_{nn'} ,
  \label{eq:Hnnkbar}
\end{equation}
in which the matrix $\bar{D}^{ij}_{nn'}$ is adjusted to maintain
agreement with all experimental effective masses.  This constraint
does not, however, completely determine $\bar{D}^{ij}_{nn'}$.

To see this, consider applying Eq.\ (\ref{eq:D}) separately to set
$\mathcal{A}$ and to the subset $\mathcal{A}_{nn'} \subseteq
\mathcal{A}$ defined by $\mathcal{A}_{nn'} = \{ |n'' \vect{k}\rangle
\mid \min(E_n, E_{n'}) \le E_{n''} \le \max(E_n, E_{n'}) \}$.  A
comparison of the results for $\mathcal{A}$ and $\mathcal{A}_{nn'}$
gives
\begin{equation}
  D^{ij}_{nn'} (\mathcal{A}_{nn'}) = D^{ij}_{nn'} + \frac12
  \sum_{l}^{\bar{\mathcal{A}}_{nn'}} \biggl( \frac{\pi^{i}_{nl}
  \pi^{j}_{ln'}}{\omega_{nl}} + \frac{\pi^{i}_{nl}
  \pi^{j}_{ln'}}{\omega_{n' l}} \biggr) ,
  \label{eq:DA01}
\end{equation}
where $D^{ij}_{nn'} \equiv D^{ij}_{nn'} (\mathcal{A})$ and
$\bar{\mathcal{A}}_{nn'} = \mathcal{A} \setminus \mathcal{A}_{nn'}$ is
the complement of $\mathcal{A}_{nn'}$ in $\mathcal{A}$.  When $E_{n} =
E_{n'}$, $D^{ij}_{nn'} (\mathcal{A}_{nn'})$ is an experimentally
measurable effective-mass parameter for the subspace
$\mathcal{A}_{nn'}$.

If $\pi^{i}_{nn'}$ is treated as an adjustable parameter
($\pi^{i}_{nn'} \rightarrow \bar{\pi}^{i}_{nn'}$) and $D^{ij}_{nn'}
(\mathcal{A}_{nn'})$ is assumed to be independent of $\{ \Delta
\pi^{i}_{nn'} \}$, then $\bar{D}^{ij}_{nn'}$ must satisfy
\begin{equation}
  \bar{D}^{ij}_{nn'} \stackrel{?}{=} D^{ij}_{nn'} (\mathcal{A}_{nn'})
  - \frac12 \sum_{l}^{\bar{\mathcal{A}}_{nn'}} \biggl(
  \frac{\bar{\pi}^{i}_{nl} \bar{\pi}^{j}_{ln'}}{\omega_{nl}} +
  \frac{\bar{\pi}^{i}_{nl} \bar{\pi}^{j}_{ln'}}{\omega_{n'l}} \biggr)
  . \label{eq:DA02}
\end{equation}
However, in general it is only necessary for Eq.\ (\ref{eq:DA02}) to
be satisfied when $E_{n} = E_{n'}$.  This still leaves some freedom of
choice in the definition of $\bar{D}^{ij}_{nn'}$.

In this paper, the modified Hamiltonian (\ref{eq:Hnnkbar}) is derived
by applying a unitary transformation $e^S$ to the original Hamiltonian
(\ref{eq:H}):
\begin{equation}
  \bar{H} = e^{-S} H e^S = H + [H,S] + \frac{1}{2!} [[H,S],S] + \cdots
  ,
  \label{eq:Hbar}
\end{equation}
where $S = -S^{\dag}$ has matrix elements only within set
$\mathcal{A}$.  The generator $S$ is defined by
\begin{subequations} \label{eq:S}
  \begin{gather}
    \langle n \vect{k} | S | n' \vect{k}' \rangle = S_{nn'} (\vect{k})
    \delta_{\vect{k}\vect{k}'} , \\ S_{nn'} (\vect{k}) = k_i
    S_{nn'}^{i} + k_i k_j S_{nn'}^{ij} ,
  \end{gather}
\end{subequations}
in which the linear coefficient is
\begin{equation}
  S_{nn'}^{i} = \frac{\Delta \pi^{i}_{nn'}}{\omega_{nn'}} .
  \label{eq:S0i}
\end{equation}
If $S_{nn'}^{ij} = 0$, the change $\Delta D^{ij}_{nn'} =
\bar{D}^{ij}_{nn'} - D^{ij}_{nn'}$ is
\begin{subequations} \label{eq:delD}
\begin{equation}
  \Delta D^{ij}_{nn'} = - \sum_{l}^{\mathcal{A}} \biggl( \frac{\Delta
  \pi^{i}_{nl} \tilde{\pi}^{j}_{ln'}}{\omega_{nl}} +
  \frac{\tilde{\pi}^{i}_{nl} \Delta \pi^{j}_{ln'}}{\omega_{n'l}}
  \biggr) , \label{eq:delD_orig}
\end{equation}
in which $\tilde{\pi}^{i}_{nn'} = \pi^{i}_{nn'} + \frac12 \Delta
\pi^{i}_{nn'}$.  Note that if we choose $\Delta \pi^{i}_{nn'} =
-\pi^{i}_{nn'}$ (for $E_{n} \ne E_{n'}$), then $\bar{\pi}^{i}_{nn'} =
0$ and Eq.\ (\ref{eq:delD_orig}) just adds extra terms to the
summation in (\ref{eq:D}).  Thus, if set $\mathcal{A}$ comprises the
highest valence and lowest conduction states, one-band effective-mass
theory is given by $\Delta \pi^{i}_{nn'} = -\pi^{i}_{nn'}$, while the
Kane model is given by $\Delta \pi^{i}_{nn'} = 0$.  (See Appendix
\ref{app:matrix} for an alternative matrix formulation of this
result.)

Equation (\ref{eq:delD_orig}) can be rewritten as
\begin{multline}
  \Delta D^{ij}_{nn'} = \frac12 \sum_{l}^{\mathcal{A}} ( \pi^{i}_{nl}
  \pi^{j}_{ln'} - \bar{\pi}^{i}_{nl} \bar{\pi}^{j}_{ln'} ) \biggl(
  \frac{1}{\omega_{n l}} + \frac{1}{\omega_{n' l}} \biggr) \\ +
  \frac{\omega_{n n'}}{2} \sum_{l}^{\mathcal{A}} \frac{\Delta
  \pi^{i}_{nl} \pi^{j}_{ln'} - \pi^{i}_{nl} \Delta
  \pi^{j}_{ln'}}{\omega_{n l} \omega_{n' l}} ,
  \label{eq:delDsimple}
\end{multline}
\end{subequations}
which shows that Eq.\ (\ref{eq:DA02}) is satisfied when $E_{n} =
E_{n'}$, but not (in general) when $E_{n} \ne E_{n'}$.  However, the
degree of freedom corresponding to the coefficient $S_{nn'}^{ij}$ in
Eq.\ (\ref{eq:S}) has not yet been used.  Let
\begin{equation}
  S^{ij}_{nn'} = \frac{\delta D^{ij}_{nn'}}{\omega_{nn'}} ,
  \label{eq:dSij}
\end{equation}
in which $\delta D^{ij}_{nn'}$ has the same symmetry as $D^{ij}_{nn'}$
and vanishes when $E_n = E_{n'}$, but is otherwise arbitrary.  This
has the effect of adding $\delta D^{ij}_{nn'}$ to the value of $\Delta
D^{ij}_{nn'}$ given by Eq.\ (\ref{eq:delD}).  In this way, one can set
the parameters $\bar{D}^{ij}_{nn'}$ for $E_n \ne E_{n'}$ to {\em any}
desired value, including zero.  This is merely a reflection of the
fact that the terms $D^{ij}_{nn'}$ with $E_n \ne E_{n'}$ do not
contribute to the single-band effective-mass Hamiltonian,
\cite{Kane80} since their contributions are of order $k^3$ or higher.

As a particular example, one could choose
\begin{multline}
  \delta D^{ij}_{nn'} = \frac{1}{2} \sum_{l}^{\mathcal{A}_{nn'}} (
  \bar{\pi}^{i}_{nl} \bar{\pi}^{j}_{ln'} - \pi^{i}_{nl} \pi^{j}_{ln'})
  \biggl( \frac{1}{\omega_{nl}} + \frac{1}{\omega_{n'l}} \biggr) \\ +
  \frac{\omega_{nn'}}{2} \sum_{l}^{\mathcal{A}} \frac{\pi^{i}_{nl}
  \Delta \pi^{j}_{ln'} - \Delta \pi^{i}_{nl}
  \pi^{j}_{ln'}}{\omega_{nl} \omega_{n'l}} ,
  \label{eq:S0ij}
\end{multline}
which would bring Eq.\ (\ref{eq:delD}) into agreement with Eq.\
(\ref{eq:DA02}).  However, including these terms would make subsequent
analysis more complicated, so for simplicity the choice $\delta
D^{ij}_{nn'} = 0$ is adopted in the remainder of this paper.  This
choice makes little practical difference, since Eq.\ (\ref{eq:S0ij})
is in fact zero in the Kane model when spin-orbit coupling is
neglected in the momentum matrix, \cite{Kane57,Kane66,Kane80,PiBr66}
which is the only example treated explicitly here.

\subsection{Definition and elimination of spurious solutions}

\label{subsec:definition}

The preceding theory can now be used to define spurious solutions
precisely.  Spurious solutions are often defined as eigenstates of the
\kp\ Hamiltonian with large wave vectors, but this definition is not
completely satisfactory because spurious states are sometimes found
well inside the first Brillouin zone. \cite{Fore97} As emphasized by
Bastard, \cite{Bast88,Bast91} more important than the magnitude of the
wave vector is its {\em instability} with respect to small changes of
the Hamiltonian parameters.  The unitary transformation
(\ref{eq:Hbar}) allows such a change of parameters to be performed
even when the \kp\ Hamiltonian is calculated directly from first
principles.

Let the wave vector be $\vect{k} = \vect{k}_{\parallel} + \unit{n}
k_{\perp}$, where $\unit{n} \cdot \unit{n} = 1$, $\unit{n} \cdot
\vect{k}_{\parallel} = 0$, and $\unit{n}$ and $\vect{k}_{\parallel}$
are real.  A spurious solution is defined here as a root $k_{\perp}(E,
\vect{k}_{\parallel})$ of the secular equation
\begin{equation}
  \det[\bar{H}(\vect{k}) - E] = \sum_{l=0}^{2N} c_l (E,
\vect{k}_{\parallel}) k_{\perp}^l = 0 \label{eq:secular}
\end{equation}
that is an {\em unbounded} function of $\{ \Delta \pi^{i}_{nn'} \}$
for small $\{ \Delta \pi^{i}_{nn'} \}$ and $\vect{k}_{\parallel}$ and
for real $E$ near the energy gap.  (Here $N$ is the dimension of set
$\mathcal{A}$.)  This definition does not encompass all possible types
of spurious solutions (see, for example, those generated by
Hamiltonian matrix elements of order $k^4$ in Sec.\
\ref{sec:numerical} and Ref.\ \onlinecite{Fore07b}), but it does
include those that can be treated effectively by the present unitary
transformation. \cite{note:Sk3} This definition has the advantage of
simplifying subsequent analysis because it focuses attention on the
asymptotic properties of the secular equation at large $k_{\perp}$
rather than the general properties of the secular equation at
arbitrary $k_{\perp}$.

Within the stated limits, all coefficients $c_l$ in the secular
equation (\ref{eq:secular}) are bounded (i.e., $|c_l| \lesssim 1$ in
atomic units).  The roots $k_{\perp}(E, \vect{k}_{\parallel})$ can
therefore be unbounded only near $c_{2N} = 0$.
For a given direction $\unit{n}$, $c_{2N}$ is just the product of
eigenvalues $\bar{d}_{\nu}(\unit{n})$ $(\nu = 1, 2, \ldots, N)$ of the
matrix $\bar{D}(\unit{n}) \equiv \hat{n}_i \hat{n}_j \bar{D}^{ij}$.
Hence, as $\{ \Delta \pi^{i}_{nn'} \}$ varies, the spurious roots
$k_{\perp}^{\mathrm{sp}}$ are unbounded near the zeros of
$\bar{d}_{\nu}(\unit{n})$, disappearing at $\bar{d}_{\nu}(\unit{n}) =
0$ because the order of the secular equation is reduced.  In typical
cases (see Sec.\ \ref{subsec:twoband}), $k_{\perp}^{\mathrm{sp}}$
changes from large real to large complex values (or vice versa) in the
neighborhood of each singular point $\bar{d}_{\nu}(\unit{n}) = 0$.

Unphysical metallic behavior can be avoided by choosing $\{ \Delta
\pi^{i}_{nn'} \}$ (or in general $S$) such that the spurious roots
disappear.  As shown below, in the Pidgeon--Brown model, \cite{PiBr66}
this can be achieved for all directions $\unit{n}$ by setting the
conduction-band mass parameter $\bar{A} = 0$, which is the choice used
in Ref.\ \onlinecite{Fore97}.  This choice may not work in all models,
but one can also choose $S$ such that $\im (k_{\perp}^{\mathrm{sp}})
\ne 0$ for all $\unit{n}$ (or more precisely such that $|\im
(k_{\perp}^{\mathrm{sp}})| > k_0 > 0$, where $k_0$ is some chosen
value).  The implementation of these choices is discussed in greater
detail in Sec.\ \ref{subsec:choice}.

\subsection{Velocity}

Although the transformation (\ref{eq:Hbar}) replaces $\bm{\pi}$ with
$\bar{\bm{\pi}}$ in the Hamiltonian, it does not do so in the velocity
$\vect{v} = -i [ \vect{x}, H ]$, where $\vect{x}$ is the coordinate.
As shown in Appendix \ref{app:coordinate}, the effective velocity
$\bar{\vect{v}} = e^{-S} \vect{v} e^S$ for set $\mathcal{A}$ is given
to first order in $k$ by an expression of the form (\ref{eq:Hnknk})
with
\begin{multline}
  \bar{v}^{i}_{nn'} (\vect{k}) = \pi^{i}_{nn'} + k_i \delta_{nn'} +
  k_j \sum_{l}^{\mathcal{B}} \biggl( \frac{\pi^{j}_{nl}
  \pi^{i}_{ln'}}{\omega_{nl}} + \frac{\pi^{i}_{nl}
  \pi^{j}_{ln'}}{\omega_{n'l}} \biggr) \\ - k_j
  \sum_{l}^{\mathcal{A}} \biggl( \frac{\Delta \pi^{j}_{nl}
  \pi^{i}_{ln'}}{\omega_{nl}} + \frac{\pi^{i}_{nl} \Delta
  \pi^{j}_{ln'}}{\omega_{n'l}} \biggr) . \label{eq:vbar}
\end{multline}
This shows that $\bar{\vect{v}}_{nn'} (\vect{k}) \ne \nabla_{\vect{k}}
\bar{H}_{nn'} (\vect{k})$ even to zeroth order in $k$.  That is, the
velocity to order $k^0$ is $\bm{\pi}$, not $\bar{\bm{\pi}}$, for both
$\vect{v}$ and $\bar{\vect{v}}$.  In the special case $\Delta
\bm{\pi}_{nn'} = -\bm{\pi}_{nn'}$, Eq.\ (\ref{eq:vbar}) is equivalent
to the optical transition matrix element given in Eq.\ (19) of Ref.\
\onlinecite{Roth59}.

\subsection{Implications for parameter fitting}

\label{subsec:fit}

The above results suggest the need for a reinterpretation of prior
work on experimental fitting of \kp\ parameters.  In a model with a
complete set of $D^{ij}_{nn'}$ parameters, the empirical masses and
Land\'e $g$ factors are not sufficient to determine $H$; in fact, for
$E_{n} \ne E_{n'}$, $\pi^{i}_{nn'}$ is arbitrary.  This indeterminacy
could in principle be resolved by fitting $\vect{v}$ to measured
oscillator strengths, but this is not usually done because optical
transition rates are considered to be less reliable than resonance
frequencies.  Instead, the most common procedure is to fix a few
values of $D^{ij}_{nn'}$ by setting the contributions from
$\mathcal{B}$ to zero or some other convenient value (see, e.g.,
Refs.\ \onlinecite{PiBr66}, \onlinecite{Law71}, \onlinecite{HeWe84},
and \onlinecite{VurMeyRM01}), thereby permitting a deterministic fit
of $\pi^{i}_{nn'}$ from frequency data.

However, this procedure is nothing but the present transformation
(albeit without explicit recognition that a change of basis is
involved) with $\bar{D}^{ij}_{nn'}$ chosen for criteria other than the
elimination of spurious solutions.  The outcome of the fitting
procedure is thus $\bar{\bm{\pi}}$, not $\bm{\pi}$ (although typically
$\bar{\bm{\pi}} \approx \bm{\pi}$).  This shows that the production of
spurious gap states by many \kp\ parameter sets is not purely a matter
of experimental necessity but at least partially an artifact of
choices made in simplifying the $\bar{D}$ matrix.  Fitting
$\bar{D}^{ij}_{nn'}$ to nonparabolic effects \cite{Agg70} fails to
resolve the quandary because the $O(k^4)$ terms needed for a correct
description of nonparabolicity have been omitted.  [Including $O(k^4)$
terms in the experimental data fitting is also of no help because it
merely shifts the indeterminacy to a larger set of parameters.]  In
the absence of direct measurements of $\bm{\pi}$, it is not possible
to distinguish $\bm{\pi}$ from $\bar{\bm{\pi}}$ (i.e., to define
unambiguously the original LK basis) without using a microscopic model
to calculate some or all of the \kp\ parameters (see, e.g., Refs.\
\onlinecite{FuZu97} and \onlinecite{Jancu05}).  Of course, the results
are then only as good as the chosen model.

\section{Heterostructures}

\label{sec:heterostructures}

The next step is to extend this change of basis from bulk crystals to
heterostructures.  Here it is applied to the nonlinear response theory
of Refs.\ \onlinecite{Fore05b}, \onlinecite{Fore07a}, and
\onlinecite{Fore07b}, in which the heterostructure is treated as a
perturbation of some virtual bulk reference crystal. To first order,
the effective $\mathcal{A}$ Hamiltonian is $H = H^{(0)} + H^{(1)}$,
where the reference Hamiltonian $H^{(0)}$ is handled according to the
above methods, and the linear Hamiltonian $H^{(1)}$ is
\cite{Fore05b,Fore07b}
\begin{subequations} \label{eq:H_lin}
\begin{equation}
  \langle n \vect{k} | H^{(1)} | n' \vect{k}' \rangle = \sideset{}{'}
  \sum_{\alpha} \theta_{\alpha} (\vect{k} - \vect{k}')
  H^{\alpha}_{nn'} (\vect{k}, \vect{k}') , \label{eq:H1}
\end{equation}
where the sum covers independent values \cite{Fore05b,Fore07a} of
$\alpha$ and
\begin{multline}
  H^{\alpha}_{nn'} (\vect{k}, \vect{k}') = E^{\alpha}_{nn'} +
  \Xi^{\alpha} \delta_{nn'} \delta_{\vect{k}\vect{k}'} + k_i
  \pi^{i\alpha}_{nn'} + \pi^{\alpha i}_{nn'} k_{i}' \\ + k_i k_j
  D^{ij\alpha}_{nn'} + k_i D^{i \alpha j}_{nn'} k_j' + D^{\alpha
  ij}_{nn'} k_i' k_j' . \label{eq:Ha}
\end{multline}
\end{subequations}
Here $\theta_{\alpha}(\vect{k})$ is the Fourier transform of
$\theta_{\alpha} (\vect{R})$, which is the change in fractional weight
of atom $\alpha$ in cell $\vect{R}$ of the heterostructure relative to
the reference crystal.  The coefficients in (\ref{eq:Ha}) are defined
in Ref.\ \onlinecite{Fore07b}; they have the symmetry of site $\alpha$
in the reference crystal and satisfy hermiticity relations such as
$D^{\alpha ij}_{nn'} = (D^{ji\alpha}_{n'n})^*$.  The superscripts on
these coefficients indicate how the coordinate and momentum operators
are ordered.  For example, in the coordinate representation, the term
proportional to $D^{i \alpha j}_{nn'}$ has the BenDaniel--Duke
ordering \cite{BenDaniel66} $D^{i \alpha j}_{nn'} p_i \theta_{\alpha}
(\vect{x}) p_j$, where $\vect{p}$ is the momentum
operator. \cite{note:k_to_p2} For a bulk perturbation of the form
$\theta_{\alpha} (\vect{k} - \vect{k}') = \theta_{\alpha}
\delta_{\vect{k}\vect{k}'}$, operator ordering is irrelevant and only
the sums $\pi^{i\alpha}_{nn'} + \pi^{\alpha i}_{nn'}$ and
$D^{ij\alpha}_{nn'} + D^{i \alpha j}_{nn'} + D^{\alpha ij}_{nn'}$ can
be distinguished.

The unitary transformation (\ref{eq:Hbar}) is now applied with $S =
S^{(0)} + S^{(1)}$, where $S^{(0)}$ is the same as (\ref{eq:S}) and
$S^{(1)}$ is defined by an expression similar to (\ref{eq:H1}) with
\begin{equation}
  S^{\alpha}_{nn'} (\vect{k}, \vect{k}') = \frac{k_{i}
  \chi^{i\alpha}_{nn'} + \chi^{\alpha i}_{nn'} k_{i}'}{\omega_{nn'}}
  . \label{eq:S1}
\end{equation}
Here $\chi^{i\alpha}_{nn'}$ is only part of the change in
$\pi^{i\alpha}_{nn'}$, since
\begin{subequations} \label{eq:delPi1}
  \begin{align}
    \Delta \pi^{i\alpha}_{nn'} & = \chi^{i\alpha}_{nn'} -
    \sum_{l}^{\mathcal{A}} \frac{\Delta \pi^{i}_{nl}
    E^{\alpha}_{ln'}}{\omega_{nl}} , \\ \Delta \pi^{\alpha
    i}_{nn'} & = \chi^{\alpha i}_{nn'} - \sum_{l}^{\mathcal{A}}
    \frac{E^{\alpha}_{nl} \Delta \pi^{i}_{ln'}}{\omega_{n'l}}
    ,
  \end{align}
\end{subequations}
where $\Delta \pi^{\alpha i}_{nn'} = (\Delta \pi^{i\alpha}_{n'n})^*$.
Likewise, the changes in the linear $D$ tensor are given by
\begin{subequations} \label{eq:delD1}
\begin{align}
  \Delta D^{ij\alpha}_{nn'} & = - \sum_{l}^{\mathcal{A}} \biggl(
  \frac{\Delta \pi^{i}_{nl} \tilde{\pi}^{j\alpha}_{ln'}}{\omega_{nl}}
  + \frac{\tilde{\pi}^{i}_{nl} \chi^{j\alpha}_{ln'}}{\omega_{n'l}}
  \biggr) ,
  \label{eq:delDija} \\
  \Delta D^{\alpha ij}_{nn'} & = - \sum_{l}^{\mathcal{A}} \biggl(
  \frac{\chi^{\alpha i}_{nl} \tilde{\pi}^{j}_{ln'}}{\omega_{nl}} +
  \frac{\tilde{\pi}^{\alpha i}_{nl} \Delta
  \pi^{j}_{ln'}}{\omega_{n'l}} \biggr) ,
  \label{eq:delDaij} \\
  \begin{split}
    \Delta D^{i\alpha j}_{nn'} & = - \sum_{l}^{\mathcal{A}} \biggl(
    \frac{\Delta \pi^{i}_{nl} \tilde{\pi}^{\alpha
    j}_{ln'}}{\omega_{nl}} + \frac{\tilde{\pi}^{i}_{nl} \chi^{\alpha
    j}_{ln'}}{\omega_{n'l}} \biggr) \\ & \phantom{= {}} -
    \sum_{l}^{\mathcal{A}} \biggl( \frac{\chi^{i \alpha}_{nl}
    \tilde{\pi}^{j}_{ln'}}{\omega_{nl}} + \frac{\tilde{\pi}^{i
    \alpha}_{nl} \Delta \pi^{j}_{ln'}}{\omega_{n'l}} \biggr) ,
  \label{eq:delDiaj}
  \end{split}
\end{align}
\end{subequations}
where $\Delta D^{\alpha ij}_{nn'} = (\Delta D^{ji\alpha}_{n'n})^*$ and
$\Delta D^{i\alpha j}_{nn'} = (\Delta D^{j\alpha i}_{n'n})^*$.  This
system of linear equations can be solved for $\Delta \pi^{\alpha}$ as
a function of $\Delta D^{\alpha}$.  An equivalent matrix formulation
of Eqs.\ (\ref{eq:delPi1}) and (\ref{eq:delD1}) is given in Appendix
\ref{app:matrix}.

\section{The Pidgeon--Brown model}

\label{sec:PB}

\subsection{Conduction band}

\label{subsec:conduction}

As an example, consider Pidgeon and Brown's formulation of the Kane
model for a zinc-blende crystal. \cite{PiBr66} The set $\mathcal{A} =
\{ \Gamma_{6\mathrm{c}}, \Gamma_{8\mathrm{v}}, \Gamma_{7\mathrm{v}}
\}$ is defined in the tensor-product basis $\{ |S\rangle, |X\rangle,
|Y\rangle, |Z\rangle \} \otimes \{ |+\rangle, |-\rangle \}$, with
spin-orbit coupling included only to order
$k^0$. \cite{Kane57,Kane66,Kane80,PiBr66} For the bulk reference
crystal, the relevant conduction-band (CB) constants are $A =
D_{SS}^{xx}$ and $P = -i \pi_{SX}^{x}$.  From Eqs.\ (\ref{eq:DA01})
and (\ref{eq:delD}), the values of $A$ and $P$ are related to the CB
effective mass $m_{\mathrm{c}}$ by
\begin{equation}
  \frac{1}{2m_{\mathrm{c}}} = A + \frac{P^2}{\epsilon_1} = \bar{A} +
  \frac{\bar{P}^2}{\epsilon_1} , \label{eq:m_c}
\end{equation}
in which $\epsilon_n$ is the $n^{\mathrm{th}}$-order reduced energy
gap:
\begin{equation}
  \frac{1}{(\epsilon_n)^n} = \frac{2}{3 (E_{\mathrm{g}})^{n}} +
  \frac{1}{3 (E_{\mathrm{g}} + \Delta_{\mathrm{so}})^{n}} ,
\end{equation}
where $E_{\mathrm{g}} = E_{6\mathrm{c}} - E_{8\mathrm{v}}$ and
$\Delta_{\mathrm{so}} = E_{8\mathrm{v}} - E_{7\mathrm{v}}$.  The value
of $\bar{P} = -i \bar{\pi}_{SX}^{x}$ needed to obtain a desired change
$\Delta A = \bar{A} - A$ is therefore
\begin{equation}
  \bar{P}^2 = P^2 - \epsilon_1 \Delta A . \label{eq:Pbar}
\end{equation}
The selection of suitable values of $\Delta A$ and $\bar{P}$ is
discussed below in Sec.\ \ref{subsec:choice}.

For the linear response in a heterostructure, there are two
independent CB partial mass coefficients, $A^{\cdot\cdot\alpha} =
A^{\alpha\cdot\cdot}$ and $A^{\cdot\alpha\cdot}$ (where
$A^{\cdot\cdot\alpha} = D_{SS}^{xx\alpha}$, $A^{\alpha\cdot\cdot} =
D_{SS}^{\alpha xx}$, and $A^{\cdot\alpha\cdot} = D_{SS}^{x\alpha x}$),
and two independent momentum parameters, $P^{\alpha\cdot} = -i
\pi_{SX}^{\alpha x}$ and $P^{\cdot\alpha} = -i \pi_{SX}^{x\alpha}$.
Upon solving Eqs.\ (\ref{eq:delPi1}) and (\ref{eq:delD1}) for the
changes $\Delta P^{\alpha\cdot}$ and $\Delta P^{\cdot\alpha}$ needed
to obtain desired values of $\Delta A^{\cdot\cdot\alpha}$ and $\Delta
A^{\cdot\alpha\cdot}$, one finds
\begin{subequations} \label{eq:delPa}
  \begin{align}
    \Delta P^{\alpha\cdot} & = -\frac{\epsilon_1 \Delta
    A^{\alpha\cdot\cdot}}{\bar{P}} - \frac{P^{\alpha\cdot} \Delta
    P}{\bar{P}} - \frac{\epsilon_1^2 \Delta A
    E_{\mathrm{c}}^{\alpha}}{2 \epsilon_2^2 \bar{P}} ,
    \label{eq:delPai} \\ \Delta P^{\cdot\alpha} & =
    -\frac{\epsilon_1 \Delta A^{\cdot\alpha\cdot}}{2\bar{P}} -
    \frac{P^{\cdot\alpha} \Delta P}{\bar{P}} + \frac{\epsilon_1^2
    \Delta A E_{\mathrm{v}}^{\alpha}}{2 \epsilon_2^2 \bar{P}} ,
    \label{eq:delPia}
  \end{align}
\end{subequations}
where $E_{\mathrm{c}}^{\alpha} = E_{SS}^{\alpha}$ and
\begin{equation}
  \frac{E_{\mathrm{v}}^{\alpha}}{\epsilon_2^2} = \frac{2
  E_{8\mathrm{v}}^{\alpha}}{3 (E_{\mathrm{g}})^2} +
  \frac{E_{7\mathrm{v}}^{\alpha}}{3 (E_{\mathrm{g}} +
  \Delta_{\mathrm{so}})^2} .
\end{equation}
If one adds (\ref{eq:delPai}) and (\ref{eq:delPia}) to obtain the
total linear change $\Delta P^{\alpha} \equiv \Delta P^{\alpha\cdot} +
\Delta P^{\cdot\alpha}$ for a bulk crystal, the result is identical to
what is obtained from linear variation of the parameters in Eq.\
(\ref{eq:Pbar}).

Equations (\ref{eq:Pbar}) and (\ref{eq:delPa}) can now be inserted
into (\ref{eq:delD1}) to determine the changes in the other mass
parameters.  The partially renormalized bulk CB Land\'e factor $g =
-i2(D_{S+,S+}^{xy} - D_{S+,S+}^{yx})$ is related to the fully
renormalized experimental value $g_{\mathrm{c}}$ by
\begin{equation}
  g_{\mathrm{c}} = g - \frac{4 P^2}{3 \delta_1} = \bar{g} - \frac{4
  \bar{P}^2}{3 \delta_1} , \label{eq:g_c}
\end{equation}
where
\begin{equation}
   \frac{1}{(\delta_n)^{n}} = \frac{1}{(E_{\mathrm{g}})^{n}} -
   \frac{1}{(E_{\mathrm{g}} + \Delta_{\mathrm{so}})^{n}} .
\end{equation}
The change $\Delta g = \bar{g} - g$ is therefore
\begin{equation}
  \Delta g = -\frac{4 \epsilon_1}{3 \delta_1} \Delta A = -\biggl(
  \frac{4 \Delta_{\mathrm{so}}}{3 E_{\mathrm{g}} + 2
  \Delta_{\mathrm{so}}} \biggr) \Delta A . \label{eq:dg}
\end{equation}
Likewise, the changes in the linear-response terms are
\begin{subequations} \label{eq:delga}
\begin{align}
  \Delta g^{\alpha\cdot\cdot} & = -\frac{4\epsilon_1}{3\delta_1} \Delta
  A^{\alpha\cdot\cdot} - \frac23 \Delta A E_{\mathrm{c}}^{\alpha}
  \biggl( \frac{\epsilon_1^2}{\delta_1 \epsilon_2^2} -
  \frac{\epsilon_1}{\delta_2^2} \biggr) , \\ \Delta
  g^{\cdot\alpha\cdot} & = -\frac{4\epsilon_1}{3\delta_1} \Delta
  A^{\cdot\alpha\cdot} + \frac43 \Delta A \biggl( \frac{\epsilon_1^2
  E_{\mathrm{v}}^{\alpha}}{\delta_1 \epsilon_2^2} - \frac{\epsilon_1
  \beta_{\mathrm{v}}^{\alpha}}{\delta_2^2} \biggr) ,
\end{align}
\end{subequations}
where
\begin{equation}
  \frac{\beta_{\mathrm{v}}^{\alpha}}{\delta_2^2} =
  \frac{E_{8\mathrm{v}}^{\alpha}}{(E_{\mathrm{g}})^2} -
  \frac{E_{7\mathrm{v}}^{\alpha}}{(E_{\mathrm{g}} +
  \Delta_{\mathrm{so}})^2} .
\end{equation}
Note that $g^{\cdot\alpha\cdot}$ is also the linear contribution to
the CB Rashba coefficient. \cite{Fore07b}

When spin-orbit coupling is neglected in the remote $\mathcal{B}$
states, \cite{Kane57,Kane66,Kane80,PiBr66} one has simply $g =
2$. \cite{PiBr66} In the original paper of Pidgeon and Brown,
\cite{PiBr66} the value of $A$ was found to have little effect on the
calculated energy levels; therefore, it was treated as an adjustable
parameter ($A \rightarrow \bar{A}$, $P \rightarrow \bar{P}$), with
$\bar{A} = \frac12$ chosen for simplicity.  The Land\'e factor,
however, was held fixed at $\bar{g} = 2$.  The present results show
that the assumption $\Delta g = 0$ must be regarded as an
approximation because it cannot be reduced to a unitary
transformation.

According to Eq.\ (\ref{eq:dg}), $\Delta g$ will be negligible in
comparison to $\Delta A$ if the spin-orbit coupling is small
($\Delta_{\mathrm{so}} \ll E_{\mathrm{g}}$).  However, the
Pidgeon--Brown model is often used in cases where
$\Delta_{\mathrm{so}} \sim E_{\mathrm{g}}$ or even
$\Delta_{\mathrm{so}} \gg E_{\mathrm{g}}$.  In such cases, setting
$\Delta g = 0$ is no more justifiable than setting $\Delta A = 0$ when
$\Delta P \ne 0$.  But this problem is easily resolved by using the
value (\ref{eq:dg}) for $\Delta g$ (assuming, of course, that the
correct original value of $A$ is known).

The Kane interband parameter $B = D_{SZ}^{xy} + D_{SZ}^{yx}$ is
neglected in the Pidgeon--Brown model \cite{PiBr66} because it
corresponds to $O(k^3)$ terms in the single-band Hamiltonian.  The
value of $B$ is not affected by the linear term $S^i_{nn'}$ in the
generator (\ref{eq:S}) (i.e., $\Delta B = 0$) because $B$ does not
depend on $P$.  Therefore, neglecting $B$ is a consistent
approximation in the sense that $B = 0$ implies $\bar{B} = 0$.
Alternatively, one could choose $\delta D^{ij}_{nn'}$ in Eq.\
(\ref{eq:dSij}) to satisfy $\delta B = -B$, thus obtaining $\bar{B} =
0$ even when $B \ne 0$.

\subsection{Valence band}

\label{subsec:valence}

\subsubsection{Zero spin-orbit coupling}

For the valence band, consider first the case without spin.  There are
four independent $\Gamma_{15\mathrm{v}}$ parameters: \cite{Lutt56} $L
= D_{XX}^{xx}$, $M = D_{XX}^{yy}$, $N = D_{XY}^{xy} + D_{XY}^{yx}$,
and $K = D_{XY}^{xy} - D_{XY}^{yx}$.  From Eqs.\ (\ref{eq:DA01}) and
(\ref{eq:delD}) we have
\begin{subequations} \label{eq:LMNK0}
\begin{align}
  L^{0} & = L - P^2 / E_{\mathrm{g}}' && = \bar{L} - \bar{P}^2 /
  E_{\mathrm{g}}' , \label{eq:L0} \\
  M^{0} & = M && = \bar{M} , \\
  N^{0} & = N - P^2 / E_{\mathrm{g}}' && = \bar{N} - \bar{P}^2 /
  E_{\mathrm{g}}' , \\
  K^{0} & = K - P^2 / E_{\mathrm{g}}' && = \bar{K} - \bar{P}^2 /
  E_{\mathrm{g}}' ,
\end{align}
\end{subequations}
where [see Eq.\ (\ref{eq:DA01})] $L^{0} \equiv L(\mathcal{A}_0)$ is
the parameter $L$ evaluated for the subset $\mathcal{A}_{0} = \{
\Gamma_{15\mathrm{v}} \}$, and $E_{\mathrm{g}}' \equiv E_{1\mathrm{c}}
- E_{15\mathrm{v}} = E_{\mathrm{g}} + \frac13 \Delta_{\mathrm{so}}$ is
the energy gap in the absence of spin-orbit splitting.  Under these
conditions $\epsilon_1 = E_{\mathrm{g}}'$, so the bulk changes are
simply
\begin{align}
  \Delta L & = \Delta N = \Delta K = - \Delta A , & \Delta M & = 0
  . \label{eq:delL}
\end{align}
Likewise, for the linear response, $\Delta M^{\alpha\cdot\cdot} =
\Delta M^{\cdot\alpha\cdot} = 0$ and
\begin{subequations} \label{eq:delLa}
  \begin{align}
    \Delta L^{\alpha\cdot\cdot} & = \Delta N^{\alpha\cdot\cdot} = \Delta
    K^{\alpha\cdot\cdot} = -\tfrac12 \Delta A^{\cdot\alpha\cdot} , \\
    \Delta L^{\cdot\alpha\cdot} & = \Delta N^{\cdot\alpha\cdot} = \Delta
    K^{\cdot\alpha\cdot} = -2 \Delta A^{\alpha\cdot\cdot} .
  \end{align}
\end{subequations}
Again, the total bulk linear variation $\Delta L^{\alpha} \equiv
\Delta L^{\cdot\cdot\alpha} + \Delta L^{\cdot\alpha\cdot} + \Delta
L^{\alpha\cdot\cdot}$ is consistent with (\ref{eq:delL}).  However,
the interchange of CB and VB operator orderings in (\ref{eq:delLa}) is
a new feature that was not predicted by the simple model of Ref.\
\onlinecite{Fore97} (where only the numerical value of $P$ was
changed, and all terms with the ordering $X^{\alpha\cdot\cdot}$ were
excluded \cite{note:P_disc}).

Although it vanishes in bulk, the linear VB momentum does have one
independent constant $R^{\cdot\alpha} = -i \pi^{z\alpha}_{XY}$ (with
$R^{\alpha\cdot} \equiv -i \pi^{\alpha z}_{XY} = -R^{\cdot\alpha}$).
\cite{Fore07b} This term is not affected by the change
(\ref{eq:delPi1}); i.e., $\Delta R^{\cdot\alpha} = \Delta
R^{\alpha\cdot} = 0$.

\subsubsection{Nonzero spin-orbit coupling}

In the Kane model, \cite{Kane57,Kane66,Kane80,PiBr66} spin-orbit
coupling is included to order $k^0$ by adding the perturbation
$H_{\mathrm{so}} = \tfrac13 \Delta_{\mathrm{so}} (\vect{I} \cdot
\bm{\sigma})$ to the Hamiltonian for set $\mathcal{A} = \{
\Gamma_{6\mathrm{c}}, \Gamma_{8\mathrm{v}}, \Gamma_{7\mathrm{v}} \}$,
where $\vect{I}$ is the orbital angular momentum
operator. \cite{Lutt56} When working in a basis that diagonalizes
$H_{\mathrm{so}}$, \cite{Kane57,Kane66,Kane80,PiBr66} it is convenient
to define the following linear transformation of the mass parameters:
\cite{Lutt56}
\begin{subequations} \label{eq:Lutt_transform}
  \begin{align}
    \gamma_1 & = -\tfrac23 (L + 2M) , & L & = -\tfrac12 (\gamma_1 + 4
    \gamma_2) , \\
    \gamma_2 & = -\tfrac13 (L - M) , & M & = -\tfrac12 (\gamma_1 - 2
    \gamma_2) , \\
    \gamma_3 & = -\tfrac13 N , & N & = -3 \gamma_3 , \\
    \kappa & = -\tfrac13 (K + 1) , & K & = -3 \kappa - 1 .
  \end{align}
\end{subequations}
Here $\gamma_1$, $\gamma_2$, $\gamma_3$, and $\kappa$ are the modified
Luttinger parameters introduced by Pidgeon and Brown. \cite{PiBr66}
Upon applying Eq.\ (\ref{eq:DA01}) to the subset $\mathcal{A}_0 = \{
\Gamma_{8\mathrm{v}} \}$, one finds the relations \cite{PiBr66}
\begin{subequations} \label{eq:mod_Lutt}
  \begin{align}
    \gamma_1^0 & = \gamma_1 + 2 P^2 / 3 E_{\mathrm{g}} , \\
    \gamma_2^0 & = \gamma_2 +   P^2 / 3 E_{\mathrm{g}} , \\
    \gamma_3^0 & = \gamma_3 +   P^2 / 3 E_{\mathrm{g}} , \\
    \kappa^0   & = \kappa   +   P^2 / 3 E_{\mathrm{g}} ,
  \end{align}
\end{subequations}
where $\gamma_1^0$, $\gamma_2^0$, $\gamma_3^0$, and $\kappa^0$ are the
original Luttinger parameters \cite{Lutt56} for
$\Gamma_{8\mathrm{v}}$.  The Luttinger parameter $q^0 = q$ is
neglected in the Kane model because (to leading order) it is
proportional to the spin-orbit splitting of the remote $\mathcal{B}$
states. \cite{HenSuz69} Since $q$ is independent of $P$, it is not
affected by the unitary transformation (\ref{eq:Hbar}).

It is also convenient to introduce a linear transformation of the form
(\ref{eq:Lutt_transform}) for the parameters $\gamma_1^0$,
$\gamma_2^0$, $\gamma_3^0$, and $\kappa^0$ in (\ref{eq:mod_Lutt}).
When this is done, one finds that the parameters $L^0$, $M^0$, $N^0$,
and $K^0$ are related to $L$, $M$, $N$, and $K$ by expressions similar
to those given earlier in Eq.\ (\ref{eq:LMNK0}).  The only difference
is that the $\Gamma_{15}$ energy gap $E_{\mathrm{g}}'$ is replaced
[see Eq.\ (\ref{eq:mod_Lutt})] by the $\Gamma_{8}$ gap
$E_{\mathrm{g}}$.

Now consider using Eq.\ (\ref{eq:DA01}) and (\ref{eq:delD}) to
determine how the effective-mass parameters change when the unitary
transformation (\ref{eq:Hbar}) is applied.  For the
$\Gamma_{8\mathrm{v}}$ submatrix, the results are similar to Eqs.\
(\ref{eq:LMNK0}) (with $E_{\mathrm{g}}' \rightarrow E_{\mathrm{g}}$)
and (\ref{eq:mod_Lutt}):
\begin{subequations} \label{eq:Q8}
  \begin{align}
    Q_8^0 & = Q_8 - P^2 / E_{\mathrm{g}} \\ & = \bar{Q}_8 - \bar{P}^2
    / E_{\mathrm{g}} ,
  \end{align}
\end{subequations}
where $Q$ is any member of the set $\{ L$, $N$, $K$, $-\frac32
\gamma_1$, $-3 \gamma_2$, $-3 \gamma_3$, $-3 \kappa \}$ ($M$ does not
depend on $P$).  The subscript 8 is added to emphasize that Eq.\
(\ref{eq:Q8}) holds for $\Gamma_{8\mathrm{v}}$ only.  Here $Q_8^0$ is
an original Luttinger parameter, while $Q_8 \equiv Q$ and $\bar{Q}_8$
are the modified Luttinger parameters before and after the unitary
transformation.

However, when Eqs.\ (\ref{eq:DA01}) and (\ref{eq:delD}) are applied to
the $\Gamma_{7\mathrm{v}}$ submatrix, the results are different:
\begin{subequations}
  \begin{align}
    Q_7^0 & = Q_7 - P^2 / (E_{\mathrm{g}} + \Delta_{\mathrm{so}}) \\ &
    = \bar{Q}_7 - \bar{P}^2 / (E_{\mathrm{g}} + \Delta_{\mathrm{so}})
    .
  \end{align}
\end{subequations}
In the Kane model, $Q_7 \equiv Q_8 \equiv Q$, but clearly $Q_7^0 \ne
Q_8^0$ and (when $\bar{P}^2 \ne P^2$) $\bar{Q}_7 \ne \bar{Q}_8$.  Such
differences also occur in the $\Gamma_{7\mathrm{v}} \times
\Gamma_{8\mathrm{v}}$ submatrix, where the parameters are given in
terms of the above results by $Q_{78}^0 = \frac12 (Q_7^0 + Q_8^0)$,
$Q_{78} = \frac12 (Q_7 + Q_8)$, and $\bar{Q}_{78} = \frac12 (\bar{Q}_7
+ \bar{Q}_8)$.  The changes in $Q$ for each submatrix are therefore
given by
\begin{gather}
  \Delta Q_{8} = -\frac{\epsilon_1 \Delta A}{E_{\mathrm{g}}} , \qquad
  \Delta Q_{7} = -\frac{\epsilon_1 \Delta A}{E_{\mathrm{g}} +
  \Delta_{\mathrm{so}}} , \nonumber \\ \Delta Q_{78} = \frac12 (\Delta
  Q_{7} + \Delta Q_{8}) . \label{eq:delQ78}
\end{gather}

The result $Q_7^0 \ne Q_8^0$ merely reflects that the Luttinger
parameters for $\Gamma_{7\mathrm{v}}$ are different from those for
$\Gamma_{8\mathrm{v}}$.  (This fact is sometimes used to obtain an
experimental fit for $P$.\cite{ScTH85,EpScCo87,GerHenBar93}) However,
the result $\Delta Q_7 \ne \Delta Q_8$ shows that the unitary
transformation (\ref{eq:Hbar}) does {\em not} preserve the initial
equality of the modified Luttinger parameters in the
$\Gamma_{8\mathrm{v}}$, $\Gamma_{7\mathrm{v}}$, and
$\Gamma_{7\mathrm{v}} \times \Gamma_{8\mathrm{v}}$ submatrices.

The latter result is hardly surprising, because the modified Luttinger
parameters are known to have different values in each submatrix when
spin-orbit coupling is treated exactly.
\cite{WeAgLa78,TrRoRa79,Trebin88} The inequality $\Delta Q_7 \ne
\Delta Q_8$ is therefore nothing new from a physical standpoint.
Nevertheless, it does serve to show that the standard experimental
data-fitting procedure---namely, treating $P$ as a fitting parameter
and defining the modified Luttinger parameters for all of set
$\mathcal{A}$ from the $\Gamma_{8\mathrm{v}}$ Luttinger parameters via
Eq.\ (\ref{eq:mod_Lutt})---is {\em not} equivalent to a simple unitary
transformation.  Instead, as $P$ is varied during the fitting, one
must invoke the additional approximation of replacing $\Delta Q_7$
with $\Delta Q_8$ in order to preserve equality of the modified
Luttinger parameters in all submatrices.

The assumption that the modified Luttinger parameters have the same
value in all submatrices even when $P$ is treated as a fitting
parameter will be referred to as the Pidgeon--Brown approximation
(PBA), since these authors seem to be the first to use it explicitly
\cite{PiBr66} (although this approximation is implicit in the theory
of Kane\cite{Kane57,Kane66,Kane80}).  The validity of the PBA was
studied by Boujdaria {\em et al}., \cite{Bouj01} who found that it
works well in many materials. \cite{note:Bouj01} However, it should be
noted that the error
\begin{equation}
  \Delta Q_8 - \Delta Q_7 = -\frac{\epsilon_1}{\delta_1} \Delta A =
  -\biggl( \frac{3 \Delta_{\mathrm{so}}}{3 E_{\mathrm{g}} + 2
  \Delta_{\mathrm{so}}} \biggr) \Delta A \label{eq:dQ7dQ8}
\end{equation}
in making the replacement $\Delta Q_7 \rightarrow \Delta Q_8$ is of
the same order as $\Delta g$ in Eq.\ (\ref{eq:dg}).  In Sec.\
\ref{subsec:conduction} it was argued that $\Delta g$ is not generally
negligible.  The error in $\Delta Q_7$ is negligible in the present
context not because $\Delta A$ is small (although it usually is), but
because this error affects primarily only the spin-orbit split-off
$\Gamma_{7\mathrm{v}}$ band.  This band is typically not of direct
experimental interest \cite{PiBr66} unless $\Delta_{\mathrm{so}} \ll
E_{\mathrm{g}}$, in which case the error (\ref{eq:dQ7dQ8}) is
negligible.

In a heterostructure, the linear changes are similar to the spin-zero
expressions (\ref{eq:delLa}).  In keeping with the PBA, only the
$\Gamma_{8}$ results are given here:
\begin{subequations} \label{eq:delLa2}
\begin{align}
  \Delta Q^{\alpha\cdot\cdot} & = -\frac{\epsilon_1 \Delta
  A^{\cdot\alpha\cdot}}{2 E_{\mathrm{g}}} - \frac{\epsilon_1 \Delta
  A}{2 E_{\mathrm{g}}} \biggl(
  \frac{E^{\alpha}_{8\mathrm{v}}}{E_{\mathrm{g}}} - \frac{\epsilon_1
  E^{\alpha}_{\mathrm{v}}}{\epsilon_2^2} \biggr) , \\ \Delta
  Q^{\cdot\alpha\cdot} & = -\frac{2 \epsilon_1 \Delta
  A^{\alpha\cdot\cdot}}{E_{\mathrm{g}}} + \frac{\epsilon_1 \Delta A
  E^{\alpha}_{\mathrm{c}}}{E_{\mathrm{g}}} \biggl(
  \frac{1}{E_{\mathrm{g}}} - \frac{\epsilon_1}{\epsilon_2^2} \biggr) .
\end{align}
\end{subequations}
The present method could, of course, be used without the PBA; this
possibility is discussed below in Sec.\ \ref{subsec:beyond}.

\subsection{Choice of parameters}

\label{subsec:choice}

Procedures for choosing $\bar{P}$ to avoid real spurious solutions
have not yet been specified.  In Sec.\ \ref{subsec:definition} it was
shown that spurious roots $k_{\perp}^{\mathrm{sp}}$ disappear at the
zeros of the eigenvalues $\bar{d}_{\nu}(\unit{n})$ of the matrix
$\bar{D}(\unit{n}) = \hat{n}_i \hat{n}_j \bar{D}^{ij}$.  In the PB
model, $\bar{D}$ is block diagonal: $\bar{D} = \bar{D}^{\mathrm{c}}
\oplus \bar{D}^{\mathrm{v}}$.  The eigenvalues of the CB block
$\bar{D}^{\mathrm{c}} (\unit{n})$ are independent of the direction
$\unit{n}$: $\bar{d}^{\mathrm{c}}_{\nu}(\unit{n}) = \bar{A}$ (where
$\nu = 1, 2$).  Thus, one can eliminate spurious solutions for all
$\unit{n}$ by setting $\bar{A} = 0$ or $\bar{P} =
\bar{P}_{\mathrm{c}}$, where
\begin{equation}
  \bar{P}_{\mathrm{c}}^2 = P^2 + \epsilon_1 A = \epsilon_1 / 2 m_c.
\end{equation}
This was the choice adopted in Ref.\ \onlinecite{Fore97}.  As shown
there, for typical semiconductors $|\epsilon_1 A| \ll P^2$, so
$\bar{P}_{\mathrm{c}} \approx P$ and the resulting changes in the
Hamiltonian are small. \cite{Fore97}

Other choices of $\bar{P}$ can also be used to obtain physically
meaningful results.  In the limiting case $\bar{P} = 0$ of single-band
effective-mass theory, all states within the energy gap are
evanescent.  Spurious solutions cannot be identified in this case
because the CB and VB states are completely decoupled.  For
infinitesimal $\bar{P}$, there is an infinitesimal anticrossing of the
evanescent gap states; the spurious solutions can then be identified
as the branches with $\im k_{\perp} \ne 0$ over the entire range of
energies near the band gap.  Spurious gap modes remain evanescent for
all $\unit{n}$ in the finite interval $0 < \bar{P}^2 < \bar{P}_0^2$,
where $\bar{P}_0^2 = \min (\bar{P}_{\mathrm{c}}^2,
\bar{P}_{\mathrm{v}0}^2)$, $\bar{P}_{\mathrm{v}0}^2 = \min_{\unit{n}}
\bar{P}_{\mathrm{v}}^2 (\unit{n})$, and $\bar{P}_{\mathrm{v}}^2
(\unit{n})$ is the smallest value of $\bar{P}^2$ where any
$\bar{d}^{\mathrm{v}}_{\nu}(\unit{n}) = 0$.

As shown in Appendix \ref{app:Pv0}, when the Luttinger parameters
satisfy $\gamma_3 \ge \gamma_2$ (which is true for most
semi\-con\-duc\-tors\cite{Law71,VurMeyRM01}), the constant
$\bar{P}_{\mathrm{v}0}^2$ is (in the PBA) simply
\begin{equation}
  \bar{P}_{\mathrm{v}0}^2 = P^2 - E_{\mathrm{g}} L = -L^0
  E_{\mathrm{g}} . \label{eq:Pv0}
\end{equation}
If $P^2$ lies within the interval $0 < P^2 < \bar{P}_{0}^2$---which is
the case in the numerical examples considered below---then the
spurious gap states in the original \kp\ Hamiltonian are evanescent
for all $\unit{n}$, and the Hamiltonian is physically acceptable
without any changes at all ($\Delta A = 0$, $\bar{P} = P$).  If not, a
valid alternative to setting $\bar{A} = 0$ is to choose a value of
$\bar{P}^2$ within this interval, \cite{Alfthan05} preferably near the
upper bound $\bar{P}_0^2$ in order to minimize $\Delta P$.  For
typical semiconductors both $|E_{\mathrm{g}} L| \ll P^2$ and
$|\epsilon_1 A| \ll P^2$, so the changes in the Hamiltonian are small
regardless of whether $\bar{P}_0^2$ is equal to
$\bar{P}_{\mathrm{c}}^2$ or $\bar{P}_{\mathrm{v}0}^2$.

\subsection{Beyond the Pidgeon--Brown approximation}

\label{subsec:beyond}

The PBA used here is open to the objection that it does not provide an
exact description of the mass of the spin-orbit split-off
$\Gamma_{7\mathrm{v}}$ band \cite{Yang05} (although, as discussed
above, it provides a good approximation in many cases\cite{Bouj01}).
This deficiency can be remedied by applying the present unitary
transformation to the Hamiltonian of Weiler {\em et al}.,
\cite{WeAgLa78} which includes a full set of independent parameters
for $\Gamma_{7\mathrm{v}}$.  In particular, the momentum matrix
element $P$ has different values $P_8$ and $P_7$ for the coupling of
$\Gamma_{6\mathrm{c}}$ to $\Gamma_{8\mathrm{v}}$ and
$\Gamma_{7\mathrm{v}}$, respectively.  Equations (\ref{eq:m_c}) and
(\ref{eq:g_c}) for the CB effective mass and $g$ factor must therefore
be replaced by
\begin{equation}
  A + \frac{2 P_8^2}{3 E_{\mathrm{g}}} + \frac{P_7^2}{3
  (E_{\mathrm{g}} + \Delta_{\mathrm{so}})} = \bar{A} + \frac{2
  \bar{P}_8^2}{3 E_{\mathrm{g}}} + \frac{\bar{P}_7^2}{3
  (E_{\mathrm{g}} + \Delta_{\mathrm{so}})}
\end{equation}
and
\begin{equation}
  g - \frac{4 P_8^2}{3 E_{\mathrm{g}}} + \frac{4 P_7^2}{3
  (E_{\mathrm{g}} + \Delta_{\mathrm{so}})} = \bar{g} - \frac{4
  \bar{P}_8^2}{3 E_{\mathrm{g}}} + \frac{4 \bar{P}_7^2}{3
  (E_{\mathrm{g}} + \Delta_{\mathrm{so}})} .
\end{equation}
Since there are two independent momentum parameters, one can choose
the values of $\bar{P}_8$ and $\bar{P}_7$ in order to achieve desired
changes in both $A$ and $g$:
\begin{subequations}
\begin{align}
  \bar{P}_8^2 & = P_8^2 - E_{\mathrm{g}} (\Delta A - \tfrac14 \Delta
  g) , \\ \bar{P}_7^2 & = P_7^2 - (E_{\mathrm{g}} +
  \Delta_{\mathrm{so}}) (\Delta A + \tfrac12 \Delta g) .
\end{align}
\end{subequations}
For example, one could choose $\Delta A = -A$ and $\Delta g = -g$ in
order to set the entire CB $\bar{D}$ matrix to zero.  Alternatively,
since $g$ has no effect on spurious solutions, one could choose
$\Delta g = 0$ in order to minimize the changes in the Hamiltonian.
In either case, Eq.\ (\ref{eq:delD}) can then be applied as usual to
determine the changes in the other $D$ parameters.  However, since the
Hamiltonian of Ref.\ \onlinecite{WeAgLa78} contains many parameters
that are not commonly used in \kp\ calculations, \cite{VurMeyRM01}
this procedure will not be carried out in detail here.  In this case,
it may be more convenient to calculate the Hamiltonian changes
numerically using the matrix equations given in Appendix
\ref{app:matrix}.

\subsection{Two-band model}

\label{subsec:twoband}

In the special case of a bulk crystal with no spin-orbit coupling and
$\vect{k} = (0,0,k)$, the $|X\rangle$ and $|Y\rangle$ valence states
are not coupled to the other states.  Spurious solutions in the PB
model are consequently confined to the two-dimensional basis $\{
|S\rangle, |Z\rangle \}$ with Hamiltonian
\begin{equation}
  \bar{H} (k) = \left[
    \begin{array}{cc}
      E_{\mathrm{c}} + \bar{A} k^{2} & i \bar{P} k  \\
      -i \bar{P} k & E_{\mathrm{v}} + \bar{L} k^{2}
    \end{array}
    \right] .
  \label{eq:H2}
\end{equation}
This case is of interest \cite{WhSh81,WhMaSh82,Trze88b,MeGoOR94,Szmu05b}
because it allows simple analytical calculations of the properties of
spurious solutions; it will also be used in some of the numerical work
in Sec.\ \ref{sec:numerical}.  The original Hamiltonian parameters are
assumed to satisfy the constraints $E_{\mathrm{g}}' = E_{\mathrm{c}} -
E_{\mathrm{v}} > 0$ and
\begin{equation}
  P^2 + E_{\mathrm{g}}' A > 0 , \qquad P^2 - E_{\mathrm{g}}' L > 0 ,
  \label{eq:pos}
\end{equation}
in which the first condition is equivalent to $m_{\mathrm{c}} > 0$
[see Eq.\ (\ref{eq:m_c})] and the second is equivalent to $L^0 < 0$
[see Eq.\ (\ref{eq:L0})].  It is also assumed that $P^2 > 0$ and
$\bar{P}^2 > 0$, which from Eq.\ (\ref{eq:Pbar}) requires that $\Delta
A < P^2 / E_{\mathrm{g}}'$.

The secular equation (\ref{eq:secular}) for the Hamiltonian
(\ref{eq:H2}) has the form $c_4 k^4 + c_2 k^2 + c_0 = 0$, where $c_4 =
\bar{A} \bar{L}$, $c_0 = (E_{\mathrm{c}} - E) (E_{\mathrm{v}} - E)$,
and
\begin{subequations}
\begin{align}
  c_2 & = \bar{A} (E_{\mathrm{v}} - E) + \bar{L} (E_{\mathrm{c}} - E)
   - \bar{P}^2 \\ & = A (E_{\mathrm{v}} - E) + L (E_{\mathrm{c}} - E)
   - P^2 ,
\end{align}
\end{subequations}
in which the second equality follows from Eqs.\ (\ref{eq:m_c}) and
(\ref{eq:L0}).  Hence, the coefficients $c_0$ and $c_2$ are invariant
with respect to the unitary transformation (\ref{eq:Hbar}).  If the
Hamiltonian (\ref{eq:H2}) were extended to include terms of order
$k^4$, then $c_4$ would also be invariant, \cite{note:k4} but this
would generate terms of higher order in the secular equation that are
not invariant.  In general, the highest-order coefficient in the
(finite-order) secular equation is not invariant with respect to the
unitary transformation (\ref{eq:Hbar}).

The general solution to the secular equation is $k^2 = (-c_2 \pm
\sqrt{c_2^2 - 4 c_4 c_0}) / 2 c_4$, which shows that for bounded
$c_l$, $k$ is unbounded only when $c_4 \rightarrow 0$, as discussed in
Secs.\ \ref{subsec:definition} and \ref{subsec:choice}.  The spurious
solutions have a particularly simple form when $E = E_{\mathrm{v}}$ or
$E = E_{\mathrm{c}}$:
\begin{subequations} \label{eq:k2sp}
  \begin{align}
    k_{\mathrm{sp}}^2 (E_{\mathrm{v}}) & = (P^2 - E_{\mathrm{g}}' L) /
    \bar{A}\bar{L} , \label{eq:k2v} \\ k_{\mathrm{sp}}^2
    (E_{\mathrm{c}}) & = (P^2 + E_{\mathrm{g}}' A) / \bar{A}\bar{L}
    . \label{eq:k2c}
  \end{align}
\end{subequations}
For other values of $E$, note that $c_2(E)$ is a linear function of
$E$ that interpolates between the values $c_2 (E_{\mathrm{v}}) = -(P^2
- E_{\mathrm{g}}' L)$ and $c_2 (E_{\mathrm{c}}) = -(P^2 +
E_{\mathrm{g}}' A)$.  Thus, for small $c_4$, $k_{\mathrm{sp}}^2(E)$
interpolates approximately linearly between the values (\ref{eq:k2v})
and (\ref{eq:k2c}).

From Eq.\ (\ref{eq:pos}), the numerators of Eqs.\ (\ref{eq:k2v}) and
(\ref{eq:k2c}) are both positive.  Therefore, the spurious solutions
are evanescent when $\bar{A} \bar{L} < 0$ and propagating when
$\bar{A} \bar{L} > 0$. \cite{Trze88b,MeGoOR94} Now
$\bar{A}\bar{L} = AL + \Delta A (L - A) - (\Delta A)^2$ is a quadratic
function of $\Delta A$ that has its maximum value when $\Delta A =
\frac12 (L - A)$.  According to Eq.\ (\ref{eq:pos}), this value of
$\Delta A$ satisfies the constraint $\Delta A < P^2 / E_{\mathrm{g}}'$
and is therefore permissible.  When $\Delta A = \frac12 (L - A)$,
$\bar{A} = \bar{L} = \frac12 (A + L)$ and thus
\begin{subequations} \label{eq:k2spmax}
  \begin{align}
    k_{\mathrm{sp}}^2 (E_{\mathrm{v}}) & = 4 (P^2 - E_{\mathrm{g}}' L)
    / (A + L)^2 , \label{eq:k2vmax} \\ k_{\mathrm{sp}}^2
    (E_{\mathrm{c}}) & = 4 (P^2 + E_{\mathrm{g}}' A) / (A + L)^2
    . \label{eq:k2cmax}
  \end{align}
\end{subequations}
These expressions give the {\em smallest} positive values of
$k_{\mathrm{sp}}^2 (E_{\mathrm{v,c}})$ that are possible for any
$\Delta A$ consistent with the given Hamiltonian parameters
$E_{\mathrm{g}}'$, $P$, $A$, and $L$.  They consequently represent the
``worst'' result that could be obtained from any unitary
transformation (\ref{eq:Hbar}).  For the special case $L = -A$, real
spurious solutions do not exist for any $\Delta A$, but for $L \ne
-A$, real spurious solutions always exist for some $\Delta A$.

It should be noted that the present theory provides the first rigorous
justification for the two-band model of White and Sham,
\cite{WhSh81,WhMaSh82} in which it is assumed to be possible always to
choose $\bar{A} > 0$ and $\bar{L} < 0$ and to invoke the limit
$\bar{A} \rightarrow 0$.  The assumption $\bar{L} = -\bar{A}$,
however, is not generally valid.

\section{Numerical examples}

\label{sec:numerical}

Numerical examples demonstrating the success of the $\bar{A} = 0$
method in eliminating spurious solutions have already been given in
Refs.\ \onlinecite{Fore97} and \onlinecite{Yang05}.  Since the present
PB Hamiltonian for the case $\bar{A} = 0$ is identical to that of
Ref.\ \onlinecite{Fore97} in bulk material, those examples will not be
repeated here.  The main new feature is the interface operator
ordering derived in Eqs.\ (\ref{eq:delPa}), (\ref{eq:delga}),
(\ref{eq:delLa}), and (\ref{eq:delLa2}).

To demonstrate the validity of these results, the $\Gamma_{15}$
valence \cite{note:no_cb} subband structure of
In$_{0.53}$Ga$_{0.47}$As/InP and GaAs/AlAs superlattices was
calculated in a plane-wave basis using the \textsc{abinit} code
\cite{Gonze02,Gonze05,ABINIT} with norm-conserving pseudopotentials
and the local-density approximation (LDA).  Spin-orbit coupling was
omitted, and all technical details were the same as in Ref.\
\onlinecite{Fore07a}.  These ``exact'' model calculations were
compared with the first-principles envelope-function (EF) theory of
Refs.\ \onlinecite{Fore05b} and \onlinecite{Fore07b}, which has no
fitting parameters (not even the mean energy).  As described in Sec.\
\ref{sec:heterostructures}, the EF Hamiltonian was constructed using
nonlinear response theory; for this purpose, the bulk reference
crystals were chosen to be In$_{0.765}$Ga$_{0.235}$As$_{0.5}$P$_{0.5}$
and Al$_{0.5}$Ga$_{0.5}$As.

\subsection{Material parameters}

\label{subsec:parameters}

Calculated values of the parameters in the Kane Hamiltonian are
presented in Table \ref{table:mass_bulk} for the bulk materials of
interest.
\begin{table}
\caption{\label{table:mass_bulk}Material parameters for several bulk
         compounds and their virtual-crystal averages.  The numbers in
         parentheses were obtained from linear interpolation.  The
         signs of $B$ and $P$ depend on the phase conventions chosen
         for the basis functions.  Here the coordinate origin was
         placed on an anion, with a neighboring cation at $\tfrac14 a
         (1,1,1)$; the phases were then fixed by setting $\langle
         \vect{x} | S \rangle > 0$ and $d \langle \vect{x} | X \rangle
         /dx > 0$ at $\vect{x} = \vect{0}$.}
\begin{ruledtabular}
  \begin{tabular}{ldddd}
    & \multicolumn{1}{c}{GaAs} &
    \multicolumn{2}{c}{Al$_{0.5}$Ga$_{0.5}$As} &
    \multicolumn{1}{c}{AlAs} \\ \hline
      $A$ & +0.280 & +0.345 & (+0.339) & +0.399 \\
      $B$ & -1.156 & -0.837 & (-0.777) & -0.398 \\
      $P$ & -0.560 & -0.552 & (-0.553) & -0.545 \\
      $L$ & -0.435 & -0.345 & (-0.343) & -0.250 \\
      $M$ & -1.511 & -1.326 & (-1.318) & -1.126 \\
      $N$ & -1.553 & -1.373 & (-1.366) & -1.179 \\
      $K$ & +2.507 & +2.310 & (+2.303) & +2.100 \\ \hline
    & \multicolumn{1}{c}{In$_{0.53}$Ga$_{0.47}$As} &
    \multicolumn{2}{c}{In$_{0.765}$Ga$_{0.235}$As$_{0.5}$P$_{0.5}$} &
    \multicolumn{1}{c}{InP} \\ \hline
      $A$ & +0.321 & +0.292 & (+0.299) & +0.278 \\
      $B$ & -0.986 & -0.954 & (-0.927) & -0.869 \\
      $P$ & -0.534 & -0.505 & (-0.504) & -0.474 \\
      $L$ & -0.429 & -0.413 & (-0.410) & -0.390 \\
      $M$ & -1.385 & -1.253 & (-1.252) & -1.119 \\
      $N$ & -1.423 & -1.296 & (-1.292) & -1.160 \\
      $K$ & +2.370 & +2.165 & (+2.162) & +1.953
  \end{tabular}
\end{ruledtabular}
\end{table}
This table includes values for the Kane parameter \cite{Kane66,Kane80}
$B = D_{SZ}^{xy} + D_{SZ}^{yx}$, although this term is neglected in
the PB model. \cite{PiBr66} The values in Table \ref{table:mass_bulk}
were calculated for the set $\mathcal{A} = \{ \Gamma_{1\mathrm{c}},
\Gamma_{15\mathrm{v}} \}$, whereas those in Ref.\ \onlinecite{Fore07b}
were for the single-band case $\mathcal{A} = \{ \Gamma_{15\mathrm{v}}
\}$.  The values of $L$, $N$, and $K$ in Table \ref{table:mass_bulk}
therefore differ from those in Table I of Ref.\ \onlinecite{Fore07b}
because they do not include the interaction with the
$\Gamma_{1\mathrm{c}}$ state $|S\rangle$.  Since the coupling to the
remote $\mathcal{B}$ states is weaker in the present case, the
parameters in Table \ref{table:mass_bulk} have a relatively small
variation between materials, and the variation is nearly linear.

To demonstrate the latter point, the numbers in parentheses in Table
\ref{table:mass_bulk} give the parameters that would be obtained for
the reference crystals if Vegard's law of linear interpolation were
valid.  Linear interpolation works well in all cases, with a maximum
error of 7\% in the $B$ parameter for Al$_{0.5}$Ga$_{0.5}$As.  This
small error, in conjunction with the fact that the total variation is
already a small perturbation, suggests that the linear perturbation
theory developed in Sec.\ \ref{sec:heterostructures} should be a good
approximation for the momentum and mass parameters.
(Quadratic-response contributions are included in the present
calculations, \cite{Fore05b,Fore07a,Fore07b} but only to order $k^0$.)

Values for the linear heterostructure parameters defined in Secs.\
\ref{sec:heterostructures} and \ref{sec:PB} are listed in Table
\ref{table:mass_linear}.
\begin{table}
  \caption{\label{table:mass_linear}Linear parameters in the
           $\Gamma_{1\mathrm{c}}$--$\Gamma_{15\mathrm{v}}$
           Hamiltonian.  Here RC stands for reference crystal, and the
           labels light and heavy holes refer to the bulk properties
           in the $\langle 100 \rangle$ directions.}
  \begin{ruledtabular}
    \begin{tabular}{llccc}
      RC  & & Al$_{0.5}$Ga$_{0.5}$As &
      \multicolumn{2}{c}{\mbox{In$_{0.765}$Ga$_{0.235}$As$_{0.5}$P$_{0.5}$}} \\
      $\alpha$ & & Ga & As & Ga \\ \hline
      Conduction & $A^{\alpha}$ &
                 $-0.189$ & $+0.090$ & $-0.176$ \\
                 & $A^{\cdot\alpha\cdot}$ & 
		 $+0.099$ & $+0.039$ & $+0.121$ \\
                 & $A^{\cdot\cdot\alpha}$ &
                 $-0.144$ & $+0.026$ & $-0.148$ \\
      Interband  & $B^{\alpha}$ & 
                 $-1.151$ & $-0.023$ & $-0.566$ \\
                 & $B^{\cdot\alpha\cdot}$ & 
		 $+0.369$ & $+0.104$ & $+0.329$ \\
                 & $B^{\cdot\cdot\alpha}$ &
                 $-0.367$ & $-0.162$ & $-0.132$ \\
                 & $B^{\alpha\cdot\cdot}$ &
                 $-1.154$ & $+0.035$ & $-0.763$ \\
      Momentum   & $P^{\alpha}$ &
                 $-0.019$ & $-0.060$ & $-0.009$ \\
                 & $P^{\alpha\cdot}$ & 
		 $-0.027$ & $-0.011$ & $-0.013$ \\
                 & $P^{\cdot\alpha}$ &
                 $+0.008$ & $-0.048$ & $+0.004$ \\
      Light hole & $L^{\alpha}$ &
                 $-0.128$ & $-0.037$ & $+0.086$ \\
                 & $L^{\cdot\alpha\cdot}$ & 
		 $+0.058$ & $-0.081$ & $+0.103$ \\
                 & $L^{\cdot\cdot\alpha}$ &
                 $-0.093$ & $+0.022$ & $-0.009$ \\
      Heavy hole & $M^{\alpha}$ &
                 $-0.387$ & $-0.329$ & $+0.130$ \\
                 & $M^{\cdot\alpha\cdot}$ &
                 $-0.039$ & $-0.109$ & $+0.093$ \\
                 & $M^{\cdot\cdot\alpha}$ &
                 $-0.174$ & $-0.110$ & $+0.018$ \\
      $k^2$ mixing & $N^{\alpha}$ &
                 $-0.319$ & $-0.300$ & $+0.167$ \\
                 & $N^{\cdot\alpha\cdot}$ &
                 $-0.136$ & $-0.042$ & $+0.025$ \\
                 & $N^{\cdot\cdot\alpha}$ &
                 $-0.091$ & $-0.129$ & $+0.071$ \\
      Land\'e    & $K^{\alpha}$ & 
                 $+0.464$ & $+0.487$ & $-0.055$ \\
      Rashba     & $K^{\cdot\alpha\cdot}$ & 
                 $+0.034$ & $+0.043$ & $+0.021$ \\
                 & $K^{\alpha\cdot\cdot}$ & 
                 $+0.215$ & $+0.222$ & $-0.038$ \\
      $\delta$ mixing & $R^{\cdot\alpha}$ & 
                  $-0.028$ & $-0.017$ & $-0.038$
    \end{tabular}
  \end{ruledtabular}
\end{table}
Here $A^{\alpha}$ denotes the total bulk value $A^{\alpha} \equiv
A^{\cdot\cdot\alpha} + A^{\cdot\alpha\cdot} + A^{\alpha\cdot\cdot} =
A^{\cdot\alpha\cdot} + 2 A^{\alpha\cdot\cdot}$.  The only other
quantities not yet defined are the $B$ parameters
$B^{\cdot\alpha\cdot} = D_{SZ}^{x\alpha y} + D_{SZ}^{y\alpha x}$,
$B^{\cdot\cdot\alpha} = D_{SZ}^{xy\alpha} + D_{SZ}^{yx\alpha}$, and
$B^{\alpha\cdot\cdot} = D_{SZ}^{\alpha xy} + D_{SZ}^{\alpha yx}$.  In
this case $B^{\alpha\cdot\cdot} \ne B^{\cdot\cdot\alpha}$, so (unlike
the other $D$ terms) there are three independent linear parameters for
$B$.  The $M$ and $R$ values in Table \ref{table:mass_linear} are the
same as those in Table III of Ref.\ \onlinecite{Fore07b}, but the
other values are different.

The operator ordering given by the parameters in Table
\ref{table:mass_linear} does not seem to follow any simple general
rules beyond the observation that $|P^{\alpha\cdot}| >
|P^{\cdot\alpha}|$ for cation perturbations and $|P^{\cdot\alpha}| >
|P^{\alpha\cdot}|$ for anion perturbations.  In particular, the
BenDaniel--Duke approximation, \cite{BenDaniel66,Bast88,Bast91} in
which mass terms of the form $A^{\cdot\alpha\cdot}$,
$B^{\cdot\alpha\cdot}$, etc., are assumed to be dominant, is clearly
not valid in most cases.  (See Ref.\ \onlinecite{Fore07b} for further
discussion of this point.)  However, since the linear changes are also
small in most cases, the particular choice of operator ordering in the
present multiband model is not as important as it would be in a
single-band model.

A comparison of the parameters in Tables \ref{table:mass_bulk} and
\ref{table:mass_linear} would seem to indicate some inconsistency in
the calculation.  For example, since the difference in Ga content
between GaAs and AlAs is just 1, the linear bulk values $A^{\alpha}$,
$B^{\alpha}$, etc., from Table \ref{table:mass_linear} should be
(approximately) numerically equal to the difference in the
corresponding bulk constants of GaAs and AlAs from Table
\ref{table:mass_bulk} (assuming that the variation is in fact linear,
as suggested by the discussion of Table \ref{table:mass_bulk} above).
However, $A(\mathrm{GaAs}) - A(\mathrm{AlAs}) = -0.119$, whereas
$A^{\alpha=\mathrm{Ga}} = -0.189$.  The error of $-0.069$ in the value
predicted by $A^{\alpha}$ is much larger than the error of $-0.006$ in
the linear interpolation for $A$ shown in Table \ref{table:mass_bulk}.

The reason for the discrepancy is the different methods used to
eliminate interband coupling in the two cases.  The bulk parameters in
Table \ref{table:mass_bulk} were calculated by first diagonalizing the
entire ($\mathcal{A} + \mathcal{B}$) Hamiltonian at $\vect{k} =
\vect{0}$ exactly, and then using perturbation theory to eliminate the
\kp\ coupling between $\mathcal{A}$ and $\mathcal{B}$.  However, in
the linear-response theory of Refs.\ \onlinecite{Fore05b} and
\onlinecite{Fore07b}, the $\vect{k}$-independent heterostructure
perturbation and the \kp\ terms are all block-diagonalized together
using a single unitary transformation. \cite{Leib75,Leib77} Since the
heterostructure perturbation $X$ and the \kp\ perturbation $Y$ do not
commute, we have $e^{S(X+Y)} \ne e^{S(X)} e^{S(Y)}$, and the two
unitary transformations yield different bulk Hamiltonian matrices for
set $\mathcal{A}$.  But the difference is merely a
$\vect{k}$-dependent unitary transformation of the form defined
previously in Eqs.\ (\ref{eq:Hbar}) and (\ref{eq:S}).

To demonstrate this, the ``errors'' in the predictions obtained from
Table \ref{table:mass_linear} for the differences in $A$, $L$, $M$,
$N$, and $K$ between materials A and B (e.g., GaAs and AlAs) were
calculated from expressions of the form
\begin{equation}
  \Delta L = \sideset{}{'} \sum_{\alpha} [\theta_{\alpha}(\mathrm{A})
  - \theta_{\alpha}(\mathrm{B})] L^{\alpha} - [L(\mathrm{A}) -
  L(\mathrm{B})] .
\end{equation}
The results are shown in Table \ref{table:linear_error}.
\begin{table}
\caption{\label{table:linear_error}Error in linear-response prediction
         of the difference in effective-mass parameters for bulk
         materials A and B.  These values should satisfy Eq.\
         (\ref{eq:delL}) if the ``error'' is not really an error but
         arises only from a unitary transformation.}
\begin{ruledtabular}
  \begin{tabular}{ldd}
    A/B & \multicolumn{1}{c}{GaAs/AlAs} &
    \multicolumn{1}{c}{In$_{0.53}$Ga$_{0.47}$As/InP} \\ \hline
      $\Delta A$ & -0.069 & -0.036 \\
      $\Delta L$ & +0.056 & +0.043 \\
      $\Delta M$ & -0.001 & -0.002 \\
      $\Delta N$ & +0.056 & +0.041 \\
      $\Delta K$ & +0.057 & +0.045
  \end{tabular}
\end{ruledtabular}
\end{table}
If the discrepancy is really due to a unitary transformation of the
form (\ref{eq:Hbar}), these errors should obey the relations given
previously in Eq.\ (\ref{eq:delL}).  These relations are clearly not
satisfied exactly, but the deviation from a pure unitary
transformation, if divided equally between conduction and valence
bands, amounts to only about 0.007 for GaAs/AlAs and 0.004 for
In$_{0.53}$Ga$_{0.47}$As/InP.  This is just the magnitude of the
linear interpolation error for these parameters shown in Table
\ref{table:mass_bulk}.

Hence, a careful examination of the material parameters shows that a
linear approximation should work very well for the effective-mass and
momentum terms.  However, it should be noted that the perturbation
theory of Refs.\ \onlinecite{Fore05b} and \onlinecite{Fore07b} yields
a \kp\ Hamiltonian that already includes a unitary transformation of
the form (\ref{eq:Hbar}) relative to the conventional Kane form of the
\kp\ Hamiltonian.  (As discussed in Sec.\ \ref{subsec:fit}, this is
also the case for most empirical \kp\ data sets found in the
literature; the difference here is that in the present theory the
effect of this transformation is {\em known} and has already been
accounted for in the operator ordering for heterostructures.)

\subsection{Valence subband structure}

\label{subsec:subbands}

As a direct test of the present theory, the $\Gamma_{15}$ va\-lence
\cite{note:no_cb} subband structure was calculated numerically for
In$_{0.53}$Ga$_{0.47}$As/InP and GaAs/AlAs superlattices in the LDA
model system described above. \cite{Fore07a,Fore07b} The
transformation (\ref{eq:Hbar}) was applied to the set $\mathcal{A} =
\{ \Gamma_{1\mathrm{c}}, \Gamma_{15\mathrm{v}} \}$ with $\bar{A} =
\lambda A$ for the reference crystal and $\bar{A}^{\alpha\cdot\cdot} =
\delta_{\lambda,1} A^{\alpha\cdot\cdot}$, $\bar{A}^{\cdot\alpha\cdot}
= \delta_{\lambda,1} A^{\cdot\alpha\cdot}$ for the linear response,
where $\lambda$ is a real parameter.  Choosing $\lambda = 1$ gives no
transformation at all, but any value $\lambda \ne 1$ modifies the bulk
value of $A$ and (for simplicity) sets the linear position dependence
of $\bar{A}$ to zero.

To determine the effect of different choices of $\lambda$, recall from
Sec.\ \ref{subsec:twoband} that the spurious solutions for $\vect{k}
\parallel \langle 100 \rangle$ are evanescent when $\bar{A} \bar{L} <
0$ and propagating when $\bar{A} \bar{L} > 0$.  From Eq.\
(\ref{eq:delL}) we have $\Delta L = -\Delta A$, hence $\bar{L} = L -
(\lambda - 1) A$.  Thus, $\bar{A}$ changes sign at $\lambda = 0$,
whereas $\bar{L}$ changes sign at $\lambda = 1 + L / A$.  Putting in
the values of $A$ and $L$ for the reference crystals in Table
\ref{table:mass_linear}, one finds that for $\lambda = 1$, the
spurious solutions for In$_{0.765}$Ga$_{0.235}$As$_{0.5}$P$_{0.5}$ are
evanescent, but values of $\lambda$ in the range $-0.415 < \lambda <
0$ yield spurious propagating modes.  However, for
Al$_{0.5}$Ga$_{0.5}$As, where $L \approx -A$, spurious propagating
modes occur only in the narrow region $0 < \lambda < 0.002$.

To obtain the most rigorous test of the present theory, one can seek
out the ``worst case'' value of $\lambda$ that gives real spurious
wave vectors with the smallest magnitude.  As shown in Sec.\
\ref{subsec:twoband}, this case corresponds to $\bar{A} = \bar{L} =
\frac12 (A + L)$ or $\lambda = \lambda_0 = \frac12 (1 + L / A)$, which
is halfway between the sign changes for $\bar{A}$ and $\bar{L}$.
Since $L \approx -A$ for Al$_{0.5}$Ga$_{0.5}$As, Eq.\
(\ref{eq:k2spmax}) shows that even the ``worst case'' real spurious
solutions in this material will have extremely large wave vectors.
Therefore, in what follows, only the In$_{0.53}$Ga$_{0.47}$As/InP
material system is studied in detail, as this provides a more
stringent test.  In this system, the
In$_{0.765}$Ga$_{0.235}$As$_{0.5}$P$_{0.5}$ reference crystal has
$\lambda_0 = -0.208$.

\begin{figure}
  \includegraphics[width=8.5cm,clip]{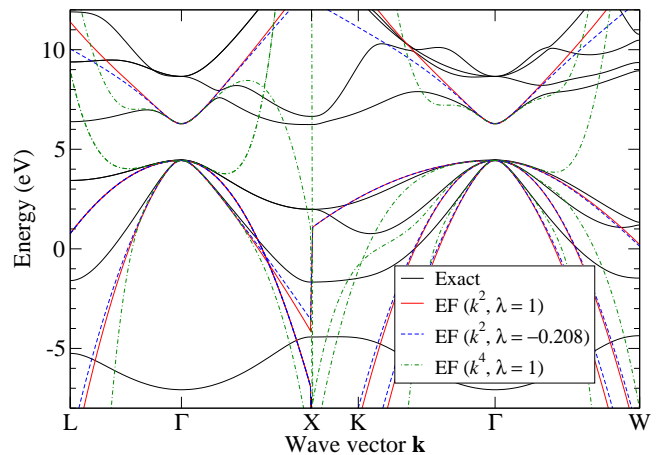}
  \caption{\label{fig:InP_bulk_bs} (Color online) Energy band
           structure of the bulk
           In$_{0.765}$Ga$_{0.235}$As$_{0.5}$P$_{0.5}$ reference
           crystal: comparison of exact calculation with 4-state \kp\
           models.}
\end{figure}

The energy band structure for
In$_{0.765}$Ga$_{0.235}$As$_{0.5}$P$_{0.5}$ is shown in Fig.\
\ref{fig:InP_bulk_bs}, which compares the ``exact'' solutions of the
model Hamiltonian with various \kp\ models.
The \kp\ calculations for $\lambda = 1$ and $\lambda = -0.208$ are
very similar for small $k$, but are visibly different for $\vect{k}$
near the Brillouin zone boundary.  The real spurious solutions for
$\lambda = -0.208$ and $\vect{k} \parallel \langle 100 \rangle$ occur
at $k \simeq \pm 15 (2\pi/a)$, where $a$ is the cubic lattice
constant.  Also shown in Fig.\ \ref{fig:InP_bulk_bs} are the results
when the \kp\ Hamiltonian is extended \cite{Fore05b,Fore07b} to
include terms of order $k^3$ and $k^4$; this case has more obvious
spurious solutions that occur well inside the Brillouin zone.

The valence subband structure of a (001)
(In$_{0.53}$Ga$_{0.47}$As)$_{24}$(InP)$_{24}$ superlattice was
calculated for a series of $O(k^2)$ EF models with $\lambda = 1$, 0.5,
0, $-0.208$, $-0.5$, and $-1$.  These calculations were performed in
momentum space \cite{GerHenBar93} with a basis containing 25 EF plane
waves (corresponding to a plane-wave cutoff at half the distance to
the bulk $X$ point).  Since the real spurious solutions occur at $|k|
\gtrsim 15 (2\pi/a)$, such a cutoff is sufficient to filter out the
spurious modes \cite{WiRo93,Fore97,Yang05} for any value of $\lambda$.

\begin{figure}
  \includegraphics[width=8.5cm,clip]{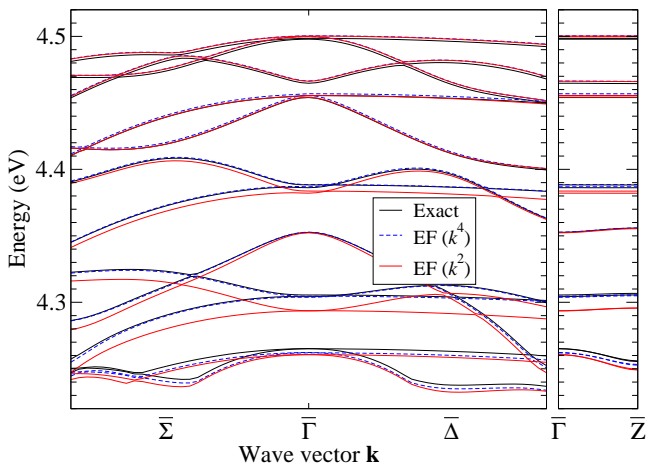}
  \caption{\label{fig:InP_slat_bs_4} (Color online) $\Gamma_{15}$
  valence subband structure of a (001)
  (In$_{0.53}$Ga$_{0.47}$As)$_{24}$(InP)$_{24}$ superlattice.}
\end{figure}

The results of these calculations are shown in Fig.\
\ref{fig:InP_slat_bs_4}.
The entire range $-1 \le \lambda \le 1$ is designated by the single
label EF($k^2$), since these values cannot be distinguished at this
scale---they differ by no more than 0.1 meV for the top five subbands
and by no more than 0.3 meV for any of the 12 subbands shown.  The
agreement with the exact calculations is excellent for the top five
subbands (with a mean error in each subband of less than 1.8 meV), but
it begins to deteriorate for energies more than 100 meV below the band
edge.  This discrepancy is due primarily to the neglect of terms of
order $k^4$ in the bulk reference Hamiltonian.  When these are
included \cite{Fore07b} [see curves labeled EF($k^4$)], the agreement
is much improved, with a maximum mean error of 3 meV for the top 12
subbands.  For the $O(k^4)$ calculations, the number of plane waves
was reduced to 17 (i.e., one-third of the Brillouin zone) in order to
avoid problems from the real spurious solutions in Fig.\
\ref{fig:InP_bulk_bs}.

The good agreement shown in Fig.\ \ref{fig:InP_slat_bs_4} confirms the
validity of both the operator ordering derived here and the
linear-response approximation used for $\pi^{i}$ and $D^{ij}$ in the
multiband EF Hamiltonian.  (Quadratic-response terms were included
only in the potential energy.\cite{Fore05b,Fore07a,Fore07b}) Note that
the 0.3 meV variation for $-1 \le \lambda \le 1$ is an order of
magnitude smaller than the 5 meV variation shown in Fig.\ 2 of Ref.\
\onlinecite{Fore97}, which did not account for changes in operator
ordering.

\section{Discussion and conclusions}

\label{sec:conclusions}

In conclusion, the unitary transformation (\ref{eq:Hbar}) eliminates
spurious solutions in the Kane model with no approximation beyond the
limitation to second-order differential operators.  A comparison of
the derived operator ordering with density-functional calculations of
the valence subband structure shows very good agreement.

This good agreement was obtained using a linear-response approximation
for the material dependence of the partially renormalized multiband
effective-mass and momentum parameters, with quadratic ``bowing''
terms included only for the band-edge energies.  As shown in Sec.\
\ref{subsec:parameters}, such an approximation is justified in a
multiband Hamiltonian because (unlike the single-band
case\cite{Fore07b}) the corrections due to renormalization from the
remote $\mathcal{B}$ states have a relatively weak variation between
materials.  Indeed, the present description of material properties is
almost identical to the interpolation scheme for the conduction-band
mass of ternary alloys recommended on page 5837 of Vurgaftman {\em et
al}.\ \cite{VurMeyRM01} (although it should be noted that this scheme
was not used uniformly for all effective-mass parameters in Ref.\
\onlinecite{VurMeyRM01}).  The only difference is that the present
work treats $P$ as linear, whereas Vurgaftman {\em et al}.\ treat
$P^2$ as linear.\cite{VurMeyRM01}

It should be noted that the linear Hamiltonian (\ref{eq:H_lin}) is
expressed in an atomistic form, as the superposition of responses to
individual atomic perturbations.  This is the simplest and most
natural way of expressing the results of linear response theory.  Such
a description may be unfamiliar to many readers since most
envelope-function models are formulated in terms of bulk compounds
rather than atoms.  However, as shown in Sec.\ VII~A of Ref.\
\onlinecite{Fore05b}, a traditional bulk-crystal description can be
obtained from the present atomistic formulation via a straightforward
linear transformation of variables (bearing in mind that the ``bulk''
compounds for the no-common-atom In$_{0.53}$Ga$_{0.47}$As/InP material
system must include not just In$_{0.53}$Ga$_{0.47}$As and InP but also
the interface materials InAs and In$_{0.53}$Ga$_{0.47}$P).
Nevertheless, there are advantages to becoming familiar with both
points of view, since the atomistic perspective is simpler and better
suited for the description of complex nanostructures.

Most envelope-function models are also expressed in a general form
that allows the inclusion of arbitrary nonlinear material dependence
in the effective-mass and momentum terms.  However, the ability to
include nonlinear terms does not necessarily imply greater accuracy,
since the present results show that the operator ordering used in most
envelope-function models is not correct even to linear order.  The
possibility of applying the present unitary transformation to a more
general phenomenological Hamiltonian with arbitrary nonlinear material
dependence was examined during the development of this paper, but
since the interface structure of the resulting theory is much more
complicated than the linear theory, it was not considered worthwhile
to publish the nonlinear results.  The linear approach has the
advantage of providing simple analytical expressions for precisely
those terms that are of greatest importance in a multiband
envelope-function theory.

For practical problems, a full implementation of the operator ordering
derived here would require the knowledge of many parameters that have
not been measured experimentally and cannot yet be predicted
accurately from first principles.  Therefore, in the near future, any
practical application of the theory based purely on existing empirical
data will require the use of some approximations.  This point is
underscored by the results obtained in Sec.\ \ref{subsec:fit}, which
show that the bulk \kp\ parameters generated by typical experimental
data-fitting procedures already include a unitary transformation---of
unknown magnitude---of the type defined here.  The uncertainty would
seem to be greatest for the convenient tabulations in review articles
\cite{VurMeyRM01} of parameters compiled from many sources.

Given such uncertainty in the existing experimental data, it is
reasonable to base short-term applications of the present theory on
the criterion of simplicity rather than theoretical rigor.  If an
unknown bulk unitary transformation is already present in the
empirical parameters, the original LK basis cannot be defined
experimentally, and it is not possible to make any definite statements
about operator ordering in heterostructures.  Therefore, one might as
well choose something simple, such as the conventional BenDaniel--Duke
operator ordering.  For simplicity, one can apply this operator
ordering {\em after} a bulk unitary transformation has been used to
eliminate real spurious solutions, as in the heuristic model of Ref.\
\onlinecite{Fore97}.

Two choices of unitary transformations in heterostructures stand out
for their simplicity.  One is to set $\bar{A} = 0$ everywhere,
\cite{Fore97} which has computational advantages \cite{Wu06} because
it allows the conduction-band envelopes to be eliminated from explicit
appearance in the envelope-function equations. \cite{Fore97} The other
is to select a single value of $\bar{P}$ for the entire
heterostructure, \cite{EpScCo87,MeGoOR94} which is chosen to yield
evanescent spurious solutions in accordance with the guidelines given
in Sec.\ \ref{subsec:choice}.  Assuming that $\bar{A} \ne 0$, this
choice simplifies the interface boundary conditions (for calculations
based on the flat-band approximation) because it ensures continuity of
all envelopes. \cite{WhSh81,WhMaSh82}

The above approach is merely a quick practical fix in which the
uncertainty in experimental parameters is openly acknowledged and even
turned to advantage by selecting simple operator ordering and a
parameter set with no real spurious solutions.  The resulting errors
in operator ordering---which are probably systematic---are simply
ignored.

However, it is hoped that the present theory will also provide a
stimulus for future work in which the sources of ambiguity in our
present knowledge are steadily eliminated.  With a careful combination
of experimental data and empirically-based microscopic theory (such as
empirical pseudopotentials \cite{FuZu97} or empirical tight-binding
theory\cite{Jancu05}) it should be possible to establish for each
material whether spurious solutions in the Kane Hamiltonian are really
required by experiment or are merely an artifact of current
data-fitting procedures.  Application of the same methods to
heterostructures will provide more definitive results for the
parameters that determine operator ordering.  At the same time,
extensions of the present {\em ab initio} techniques to include
quasiparticle self-energies and projector-augmented waves should
provide more accurate predictions of parameters from first principles.
It is hoped that at some time in the near future these two lines of
investigation will converge to yield a practical \kp\ theory free from
ambiguity.

\begin{acknowledgments}
This work was supported by Hong Kong RGC Grant No.\ 600905.
\end{acknowledgments}

\appendix

\section{Matrix formulation of Hamiltonian changes}

\label{app:matrix}

Let $H_m^{(n)}$ be the contribution to the matrix $H$ that is of order
$\theta^n k^m$, where $\theta$ is the heterostructure perturbation
parameter introduced in Eq.\ (\ref{eq:H1}).  Then the changes in the
reference crystal Hamiltonian (\ref{eq:H}) due to the unitary
transformation (\ref{eq:Hbar}) are given by [cf.\ Eq.\
(\ref{eq:delD})]
\begin{align}
  \Delta H_1^{(0)} & = [ H_0^{(0)}, S_1^{(0)} ] , \\
  \Delta H_2^{(0)} & = [ \tilde{H}_1^{(0)}, S_1^{(0)} ] 
                     + [ H_0^{(0)}, S_2^{(0)} ] ,
\end{align}
while the changes in the linear Hamiltonian (\ref{eq:H_lin}) are [cf.\
Eqs.\ (\ref{eq:delPi1}) and (\ref{eq:delD1})]
\begin{align}
  \Delta H_1^{(1)} & = [ H_0^{(1)}, S_1^{(0)} ] +
                       [ H_0^{(0)}, S_1^{(1)} ] , \\
  \begin{split}
    \Delta H_2^{(1)} & = [ \tilde{H}_1^{(1)}, S_1^{(0)} ] +
                         [ \tilde{H}_1^{(0)}, S_1^{(1)} ]  \\ & {} +
                         [ H_0^{(1)}, S_2^{(0)} ] + 
                         [ H_0^{(0)}, S_2^{(1)} ] ,
  \end{split}
\end{align}
in which $\tilde{H}_1^{(n)} = H_1^{(n)} + \frac12 \Delta H_1^{(n)}$.

\section{Coordinate and velocity}

\label{app:coordinate}

This appendix examines the effect of the transformation
(\ref{eq:Hbar}) on the coordinate and velocity.  In the LK
representation, the coordinate operator inside the first Brillouin
zone is just $i \nabla_{\vect{k}} \delta_{nn'}$. \cite{LuKo55} After
the \kp\ coupling between $\mathcal{A}$ and $\mathcal{B}$ is
eliminated, the effective coordinate for $\mathcal{A}$ becomes
\begin{subequations}
\begin{equation}
  \langle n \vect{k} | \vect{x} | n' \vect{k}' \rangle =
  \vect{x}_{nn'} (\vect{k}) \delta (\vect{k} - \vect{k}') ,
\end{equation}
in which the operator $\vect{x}_{nn'} (\vect{k})$ is given to first
order in $k$ by
\begin{equation}
  \vect{x}_{nn'} (\vect{k}) = i \nabla_{\vect{k}} \delta_{nn'} +
  \frac12 \, \bm{\Omega}_{nn'} \times \vect{k} .
\end{equation}
\end{subequations}
Here $\bm{\Omega}_{nn'}$ is the Berry curvature
\cite{AdBl59,Blount62,Lax74,Berry84,SunNiu99} at $\vect{k} = \vect{0}$
for the quasi-Bloch (or transformed LK) basis:
\begin{equation}
  \bm{\Omega}_{nn'} = i \sum_{l}^{\mathcal{B}} \bm{\xi}_{nl} \times
  \bm{\xi}_{ln'} ,
\end{equation}
in which $\bm{\xi}_{nn'}$ is the crystal coordinate \cite{Adams52} or
Berry connection \cite{Berry84}
\begin{equation}
  \bm{\xi}_{nn'} = \frac{-i \bm{\pi}_{nn'}}{\omega_{nn'}} \qquad
  (E_{n} \ne E_{n'}) .
\end{equation}
The effective velocity $\vect{v} = -i [ \vect{x}, H ]$ for
$\mathcal{A}$ is therefore given (to first order in $k$) by
\begin{equation}
  \vect{v}_{nn'} (\vect{k}) = \nabla_{\vect{k}} H_{nn'} (\vect{k}) +
  \frac{i \omega_{nn'}}{2} \bm{\Omega}_{nn'} \times \vect{k}
  . \label{eq:v}
\end{equation}
Here the contribution from $\bm{\Omega}_{nn'}$ vanishes in a
single-band effective-mass model, but not in a multiband model.  This
contribution is related \cite{note:anom_vel} to the so-called
anomalous velocity \cite{AdBl59,Blount62} or Hall velocity
\cite{SunNiu99} in an external field.  $\bm{\Omega}$ is a hermitian
operator with the same symmetry as an angular momentum or a magnetic
field (i.e., $\bm{\Omega}$ is a pseudovector that is odd under time
reversal).  In a zinc-blende crystal, $\bm{\Omega}$ has $\Gamma_{25'}$
symmetry and couples the $\Gamma_{6}$ conduction band to the
$\Gamma_{8}$ valence band in the presence of spin-orbit coupling.

After the transformation $\bar{\vect{x}} = e^{-S} \vect{x} e^S$, the
effective coordinate for $\mathcal{A}$ becomes
\begin{multline}
  \bar{\vect{x}}_{nn'} (\vect{k}) = i \nabla_{\vect{k}} \delta_{nn'} +
  \Delta \bm{\xi}_{nn'} + \frac12 \, \bar{\bm{\Omega}}_{nn'} \times
  \vect{k} \\ + i \hat{\vect{x}}_j (S^{jl}_{nn'} + S^{lj}_{nn'}) k_l ,
\end{multline}
in which $\Delta \bm{\xi}_{nn'} = +i \Delta \bm{\pi}_{nn'} /
\omega_{nn'}$ and $\bar{\bm{\Omega}}_{nn'} = \bm{\Omega}_{nn'} +
\Delta \bm{\Omega}_{nn'}$, where
\begin{equation}
  \Delta \bm{\Omega}_{nn'} = i \sum_{l}^{\mathcal{A}} \Delta
  \bm{\xi}_{nl} \times \Delta \bm{\xi}_{ln'} .
\end{equation}
The transformed velocity $\bar{\vect{v}} = e^{-S} \vect{v} e^S = -i [
\bar{\vect{x}}, \bar{H} ]$ is given by Eq.\ (\ref{eq:vbar}).  In this
case, attempting to write $\bar{\vect{v}}$ in a form analogous to Eq.\
(\ref{eq:v}) yields a rather lengthy expression that is not given
here.

\section{Equation (\ref{eq:Pv0})}

\label{app:Pv0}

This appendix contains a derivation of the expression for
$\bar{P}_{\mathrm{v}0}^2 = \min_{\unit{n}} \bar{P}_{\mathrm{v}}^2
(\unit{n})$ given in Eq.\ (\ref{eq:Pv0}) of Sec.\ \ref{subsec:choice}.
Here $\bar{P}_{\mathrm{v}}^2 (\unit{n})$ is the smallest value of
$\bar{P}^2$ where any eigenvalue $\bar{d}^{\mathrm{v}}_{\nu}(\unit{n})
= 0$.  To find the value of $\bar{P}_{\mathrm{v}0}^2$ it is therefore
necessary to determine the direction $\unit{n}$ in which
$\bar{d}^{\mathrm{v}}_{\nu}(\unit{n})$ first reaches zero (for any
$\nu$ or $\unit{n}$) as $\bar{P}^2$ is increased from zero.

The problem can be simplified by noting that in the PBA, the $6 \times
6$ VB block $\bar{D}^{\mathrm{v}} (\unit{n})$ can be further reduced
to the direct sum of two $3 \times 3$ spin-zero blocks, since the mass
parameters in the PBA do not depend on spin.  The eigenvalues of these
$3 \times 3$ matrices cannot be found analytically for general
$\unit{n}$, but a useful approximate solution can be obtained from a
rotated basis \cite{Kane57} $\{ |X'\rangle, |Y'\rangle, |Z'\rangle \}$
in which $|Z'\rangle = \hat{n}_x |X\rangle + \hat{n}_y |Y\rangle +
\hat{n}_z |Z\rangle$.  In this basis, the $|Z'\rangle$ state is of
principal interest because $|X'\rangle$ and $|Y'\rangle$ are not
coupled to the CB by the \kp\ interaction.  The corresponding diagonal
matrix element of $D^{\mathrm{v}} (\unit{n})$ is
\begin{equation}
  D^{\mathrm{v}}_{Z'Z'} (\unit{n}) = L - 2 (L - M - N) (\hat{n}_y^2
  \hat{n}_z^2 + \hat{n}_z^2 \hat{n}_x^2 + \hat{n}_x^2 \hat{n}_y^2 ) .
  \label{eq:DZZa}
\end{equation}
For $\unit{n}$ in the $\langle 100 \rangle$, $\langle 110 \rangle$,
and $\langle 111 \rangle$ directions, the matrix $D^{\mathrm{v}}
(\unit{n})$ is diagonal and Eq.\ (\ref{eq:DZZa}) is an exact
eigenvalue.  For other directions, Eq.\ (\ref{eq:DZZa}) is not an exact
eigenvalue, but it does provide a useful qualitative description of
the angular dependence of the exact solution.

In typical semiconductors, the Luttinger parameters
(original\cite{Lutt56} or modified\cite{PiBr66}) satisfy $\gamma_3 >
\gamma_2$, \cite{Law71,VurMeyRM01} hence $L-M-N = 3(\gamma_3 -
\gamma_2) > 0$.  Equation (\ref{eq:DZZa}) therefore suggests that the
first eigenvalue $\bar{d}^{\mathrm{v}}_{\nu}(\unit{n})$ (for any
direction $\unit{n}$) to reach $\bar{d}^{\mathrm{v}}_{\nu}(\unit{n}) =
0$ as $\bar{P}$ increases from zero will be the eigenvalue $\bar{L}$
corresponding to a state $|Z'\rangle$ with $\unit{n} \parallel \langle
100 \rangle$.  This tentative conclusion has been confirmed by a
numerical examination of the eigenvalues
$\bar{d}^{\mathrm{v}}_{\nu}(\unit{n})$ in different directions as
$\bar{P}$ is varied.

Therefore, when $\gamma_3 \ge \gamma_2$ (as is usually the case), the
constant $\bar{P}_{\mathrm{v}0}^2$ is given by Eq.\ (\ref{eq:Pv0}).


\begin{thebibliography}{88}
\bibitem[{not({\natexlab{a}})}]{note:Fore97}
\bibinfo{note}{For readers of Ref.\ \onlinecite{Fore97} who wonder what
  ``envelope structure theory'' is, this phrase was coined by an overzealous
  editor---after final proof corrections---as an abridgment of
  ``heterostructure envelope-function theory.''}
\bibitem[{not({\natexlab{b}})}]{note:Sk3}
\bibinfo{note}{It is possible that the real spurious solutions produced by
  $O(k^4)$ terms in the multiband Hamiltonian could be eliminated by extending
  the generator $S$ in Eq.\ (\ref{eq:S}) to include terms of order $k^3$, but
  this possibility will not be considered further here.}
\bibitem[{not({\natexlab{d}})}]{note:P_disc}
\bibinfo{note}{The term $P^{\cdot\alpha}$ (but not $P^{\alpha\cdot}$) was also
  excluded in Ref.\ \onlinecite{Fore97} because of a mathematical conflict in
  the model with $\bar{A} = 0$ and piecewise-constant material parameters. This
  conflict remains, but the correct resolution is not to abandon the operator
  ordering derived from first principles. Rather, one should reject
  discontinuous parameters as physically unrealistic.}
\bibitem[{not({\natexlab{f}})}]{note:k4}
\bibinfo{note}{In Ref.\ \onlinecite{Kolokolov03}, Kolokolov {\em et al}.\ state
  that their modification of the Kane Hamiltonian has less error (measured with
  respect to the original Kane Hamiltonian) in the coefficient of $k^4$ in the
  secular equation than the model of Ref.\ \onlinecite{Fore97}. This assertion
  is meaningless, however, since the Kane Hamiltonian is not valid to order
  $k^4$. It should also be noted that the modification proposed in Ref.\
  \onlinecite{Kolokolov03} is not invariant with respect to the symmetry
  operations of the $T_d$ group.}
\bibitem[{not({\natexlab{g}})}]{note:no_cb}
\bibinfo{note}{The conduction subband structure was not considered here because
  (see Fig.\ \ref{fig:InP_bulk_bs}) the $L$ conduction valleys of the model
  system are nearly degenerate with the $\Gamma$ valley. \cite{Fore07a,Fore07b}
  Hence, a correct treatment of the conduction subbands in the model system
  would require the inclusion of intervalley mixing, which is beyond the scope
  of the Kane model. (To put this in another way, the exact conduction subbands
  of the LDA model system have ``spurious solutions'' of their own that would
  lead to unnecessary complications.)}
\bibitem[{not({\natexlab{h}})}]{note:anom_vel}
\bibinfo{note}{The relationship is not an exact equivalence because the
  quasi-Bloch basis differs from a true Bloch basis and because external fields
  are not included here.}
\end{thebibliography}

\end{document}